%% file: main.tex
\documentclass[11pt]{article}

\usepackage[utf8]{inputenc}
\usepackage[T1]{fontenc}
\usepackage{lmodern}
\usepackage[english]{babel}
\usepackage{microtype}
\usepackage{amsmath,amssymb,amsthm}
\usepackage{mathtools}
\usepackage{graphicx}
\usepackage{booktabs}
\usepackage{array}
\usepackage{makecell}
\usepackage{multirow}
\usepackage{tabularx}
\usepackage{rotating}
\usepackage{xcolor}
\usepackage{tikz}
\usepackage{pgfplots}
\pgfplotsset{compat=1.16}
\usetikzlibrary{positioning,arrows.meta,calc,fit,backgrounds,shapes.geometric,decorations.pathreplacing}
\usepackage[font=small,labelfont=bf,labelsep=period]{caption}
\usepackage[round,authoryear]{natbib}
\usepackage{enumitem}
\usepackage[a4paper,margin=2.6cm]{geometry}
\usepackage[colorlinks=true,allcolors=linkcol]{hyperref}

\definecolor{linkcol}{HTML}{1F5C73}   
\definecolor{pInk}{HTML}{1A1A1A}      
\definecolor{pTeal}{HTML}{1F5C73}     
\definecolor{pTealL}{HTML}{6FA8B8}    
\definecolor{pGold}{HTML}{B8860B}     
\definecolor{pGoldL}{HTML}{E6C66B}    
\definecolor{pMute}{HTML}{8A8A8A}     
\definecolor{pBg}{HTML}{F3F1EC}       
\definecolor{pState}{HTML}{6E4B7A}    
\definecolor{pMarket}{HTML}{8A8A8A}   
\definecolor{pCommons}{HTML}{1F5C73}  

\theoremstyle{definition}
\newtheorem{definition}{Definition}
\newtheorem{principle}{Design Principle}
\theoremstyle{remark}
\newtheorem{observation}{Observation}

\renewcommand{\arraystretch}{1.18}
\newcommand{\stack}[1]{\textsc{#1}}

\title{\bfseries Commons-Governed Artificial Intelligence:\\[2pt]
A Taxonomy of Collective Governance}

\author{
  Eduardo C. Garrido-Merch\'an\\
  \small Department of Quantitative Methods, ICADE\\
  \small Universidad Pontificia Comillas, Madrid, Spain\\
  \small \texttt{ecgarrido@comillas.edu}
}
\date{}

\begin{document}
\maketitle

\begin{abstract}
\noindent
The governance of artificial intelligence is overwhelmingly theorized through two
institutional frames. In the market frame, the data, models, and compute that
constitute the AI stack are private goods exchanged under property and contract;
in the state frame, a regulator imposes rules from above. A third possibility,
the collective and self-organized stewardship of AI-relevant resources by the
communities that produce and depend on them, remains comparatively
under-theorized, even as it proliferates in practice through data trusts and
cooperatives, federated learning consortia, public compute initiatives,
open-weight model collaborations, and community data sovereignty regimes. This
article argues that these arrangements form a coherent institutional family,
which we call commons-governed artificial intelligence, and that the analytic
vocabulary developed by Elinor Ostrom and her successors for common-pool and
knowledge commons is the right backbone for classifying them. We contribute a
two-dimensional taxonomy whose first axis is the resource layer of the AI stack
held in common, distinguishing data, compute, models, knowledge and evaluation,
and energy, and whose second axis is the governance function performed, derived
from Ostrom's design principles of boundary definition, appropriation and
provision congruence, collective choice, monitoring, graduated sanctioning,
conflict resolution, recognition of the right to organize, and nested
polycentric scaling. We populate the taxonomy by examining the published evidence
layer by layer, locate ten recurrent institutional archetypes within it, synthesize their
positions through a maturity matrix and a comparative reading against the eight
design principles, and treat the energy and sustainability of computation as a
first-class commons-governance problem rather than an externality. We close with
the tensions that constrain the project, openwashing, the compute bottleneck,
free-riding, and the tension between scale and sustainability, and with a research
and policy agenda for a polycentric AI commons. The taxonomy gives the standing
critique that ethical principles alone cannot govern AI a constructive
institutional answer.

\medskip
\noindent\textbf{Keywords:} AI governance; commons; Ostrom; common-pool resources;
data trusts; data cooperatives; federated learning; compute governance; open-weight
models; sustainable AI; polycentric governance.
\end{abstract}

\input{sec_intro}
\input{sec_commons}
\input{sec_genealogy}
\input{sec_stack}
\input{sec_taxonomy}
\input{sec_data}
\input{sec_compute}
\input{sec_models}
\input{sec_knowledge}
\input{sec_energy}
\input{sec_synthesis}
\input{sec_applications}
\input{sec_tensions}
\input{sec_agenda}
\input{sec_conclusion}

\bibliographystyle{plainnat}
\bibliography{bib/refs}

\end{document}

%% file: sec_intro.tex
\section{Introduction}
\label{sec:intro}

Artificial intelligence is now built from a stack of resources whose scale and
concentration are without precedent in the history of computing. Frontier models
are trained on corpora assembled from the textual and visual output of much of
humanity, on clusters of specialized accelerators available to only a handful of
firms and states, and at an energy cost that has become a measurable fraction of
global electricity demand \citep{iea2025energyai}. The question of how this stack
should be governed has accordingly moved from a specialist concern to a central
problem of contemporary political economy. Yet the dominant ways of posing that
question presuppose one of two institutional answers. The first is the market
answer, in which data, model weights, and compute are private goods, allocated
through property and contract, and governed by the firms that own them. The second
is the state answer, in which a public regulator imposes binding rules from above,
as in the European Union's Artificial Intelligence Act \citep{euaiact2024}, the
risk-management framework of the United States National Institute of Standards and
Technology \citep{nist2023airmf}, or the intergovernmental principles of the
Organisation for Economic Co-operation and Development \citep{oecd2019recommendation}.
Between the market and the state lies a third institutional possibility that the
governance literature has so far treated only in fragments: the collective,
self-organized stewardship of AI-relevant resources by the communities that
produce and depend on them. We call this possibility \emph{commons-governed
artificial intelligence}, and this article contributes a taxonomy of it.

The claim that there exists a coherent third way is not a rhetorical flourish but
the central empirical finding of the commons tradition in institutional analysis.
\citet{hardin1968tragedy} argued that a resource held in common is doomed to
overexploitation, because each appropriator captures the full benefit of an
additional withdrawal while bearing only a fraction of its cost, so that, in his
phrase, freedom in a commons brings ruin to all. The policy corollary, that only
privatization or external coercion can avert collapse, is precisely the dichotomy
the market and state frames inherit. The decisive rebuttal is the body of work for
which \citet{ostrom1990governing} received the Nobel Memorial Prize in Economic
Sciences. By distinguishing open-access regimes, which were Hardin's actual model,
from common-pool resources governed by bounded communities under self-determined
rules, and by documenting durable self-governing institutions across irrigation
systems, fisheries, forests, and groundwater basins, Ostrom established both
theoretically and empirically that the space between markets and states is
institutionally occupied. Her account of polycentric governance, the thesis that
complex resource systems are stewarded effectively by multiple overlapping centers
of authority rather than by a single monocentric one, gives the third way its
structure \citep{ostrom2010beyond}. The organizing wager of this article is that
the resources of the AI stack, training corpora, model weights, evaluation
benchmarks, pooled compute, and the open toolchains that bind them, exhibit the
structural features of common-pool and knowledge commons, and that the analytic
vocabulary built for those resources is the right instrument for classifying how
AI is, and might be, governed in common.

That wager is supported by a rapidly growing but fragmented practice. On the data
layer, bottom-up data trusts vest fiduciary control of pooled data in a trustee
bound to the interests of the data subjects \citep{delacroix2019bottomup}, data
cooperatives place curation and analytics under member ownership on the model of
the credit union \citep{hardjono2019datacoops}, and community regimes such as the
CARE principles for Indigenous data governance assert collective rather than merely
individual control over data \citep{carroll2020care}. On the compute layer, the
United States National AI Research Resource and the European EuroHPC initiative
pool publicly funded computational capacity for broad access
\citep{nairr2023strengthening,eurohpc2024aifactories}, and federated learning makes
it possible to train a shared model over decentralized data without pooling the raw
data itself \citep{mcmahan2017communication,kairouz2021advances}. On the model
layer, collaborations such as the BigScience workshop have produced openly licensed
large language models on publicly granted supercomputers
\citep{bigscience2022bloom}, while transparency instruments such as model cards
\citep{mitchell2019model} and the Foundation Model Transparency Index
\citep{bommasani2023foundation,bommasani2024foundation} supply the accountability
machinery a commons would require. Each of these is typically studied in isolation,
under the vocabulary of its own subfield. None of the existing maps of the
governance terrain treats them as instances of a single institutional family.

The two literatures that come closest each stop short of the category we propose.
The AI-governance literature is organized around principles, risks, and regulatory
compliance rather than around institutions. The most cited map of the normative
terrain, the meta-analysis of \citet{jobin2019global}, documents convergence on
five ethical principles across eighty-four guideline documents, and the synthesis
of \citet{floridi2018ai4people} consolidates them into a unified ethical framework;
but neither classifies the institutional forms through which such principles might
be enacted. \citet{mittelstadt2019principles} argues precisely that principles
alone cannot guarantee ethical AI, because the principlist analogy with bioethics
fails in the absence of common aims, fiduciary duties, professional histories, and
accountability mechanisms. The most comprehensive recent stocktake, the AI Risk
Repository of \citet{slattery2024airisk}, is a database and taxonomy of risks,
deliberately agnostic about institutional remedies. The data-governance literature,
by contrast, is institutionally rich but was largely written before the
foundation-model era and is organized around personal data rather than AI training
assets. Its proximate taxonomy, the four emerging models of
\citet{micheli2020emerging}, namely data-sharing pools, data cooperatives, public
data trusts, and personal data sovereignty, is classified by who holds power and to
whose benefit data flows, not by commons-governance mechanism, and it does not
treat model or compute governance as objects in their own right. The gap this
article fills sits at the intersection: there is as yet no peer-reviewed typology
that takes the commons as the organizing principle, applies it across the full AI
stack of data, compute, models, knowledge, and energy in the foundation-model era,
and uses Ostrom's design principles as its comparative axes.

We close that gap with a two-dimensional taxonomy. The first axis is the
\emph{resource layer} of the AI stack that is held in common. We distinguish five
layers, data, compute, models and weights, knowledge and evaluation, and energy,
and we argue that each satisfies the structural criteria for treatment as a
commons, in particular that each is subtractable in some governance-relevant
dimension even where its informational content is non-rival
\citep{hess2007understanding}. The second axis is the \emph{governance function}
performed by an institution over a layer, and here we adopt the eight design
principles that \citet{ostrom1990governing} identified as characteristic of
long-enduring self-governed institutions: the definition of clear boundaries around
both resource and community, the congruence of appropriation and provision rules
with local conditions, collective-choice arrangements that let those affected
modify the rules, monitoring, graduated sanctions, accessible conflict-resolution
mechanisms, minimal recognition of the right to organize, and, for larger systems,
organization in nested enterprises. The Cartesian product of the two axes yields a
classificatory space in which any commons-governed AI institution can be located,
characterized by the layers it pools and the functions it performs over each. We
populate this space by examining the published evidence layer by layer, by
identifying ten recurrent institutional archetypes, and by reading their positions
through a maturity matrix and a comparison against the eight principles.

The contribution of the article is fourfold. It establishes commons-governed AI as
a named institutional category and grounds it in the Ostromian tradition, giving
the institutional critique of principlism a constructive answer. It proposes a
taxonomy whose axes are inherited from a validated institutional grammar rather
than constructed ad hoc, and it applies the taxonomy across the entire AI stack
rather than to data alone. It treats the energy and carbon cost of computation as a
commons-governance problem on a par with the governance of data and weights, rather
than as an externality to be priced, drawing the measurement literature
\citep{strubell2019energy,patterson2021carbon,luccioni2023estimating} and the
community-energy tradition \citep{wade2025energy} into a single frame. And it sets
out the failure modes that constrain the project, including the openwashing
identified by \citet{widder2024open}, the compute bottleneck analyzed by
\citet{sastry2024computing}, and the classical risk of free-riding, together with a
research and policy agenda for a polycentric AI commons.

The remainder of the article proceeds as follows. The next section reconstructs the
commons tradition from Hardin and Ostrom through the knowledge-commons and
commons-based peer-production literatures, isolating the conceptual moves the
taxonomy depends on. The third section argues that the AI stack is a layered
commons and defines the five resource layers. The fourth section constructs the
taxonomy itself, stating its axes, its descriptive secondary dimensions, and the
method by which it was built. The five sections that follow examine the data,
compute, model, knowledge, and energy layers in turn. A synthesis section then
assembles the institutional archetypes, presents the maturity matrix and the
principle-by-principle comparison, and reads the durable combinations off the
evidence. A penultimate pair of sections treats sectoral applications and the
tensions and failure modes of the project, and a final section presents the
research and policy agenda before concluding.

%% file: sec_commons.tex
\section{The Commons Tradition and Its Extension to Knowledge}
\label{sec:commons}

The conceptual backbone of the taxonomy is the theory of the commons and its
successive extensions from natural resources to knowledge, software, and data. We
reconstruct that tradition here only to the depth the taxonomy requires, isolating
the four moves on which the rest of the article depends: the distinction between
open access and a governed commons, the design principles that characterize durable
self-governance, the reframing of knowledge as a commons, and the account of
commons-based peer production that explains how many AI commons are produced rather
than merely held.

The first move is the separation of open access from the commons proper. The
intuition that shared resources collapse under self-interest, formalized by
\citet{hardin1968tragedy}, models a pasture to which access is unrestricted and in
which no rule constrains appropriation. Under those conditions the marginal benefit
of an additional animal accrues to the individual herder while the marginal cost of
degradation is socialized, and the dominant strategy is to overgraze. The error
that the commons tradition exists to correct is the identification of this
open-access regime with all forms of shared ownership. \citet{ostrom1990governing}
showed that the resources Hardin had in mind are in practice rarely open access;
they are common-pool resources governed by bounded communities through
self-determined institutions, and the empirical record of irrigation districts,
inshore fisheries, alpine pastures, and forest commons contains many that have
endured for centuries. A common-pool resource in this technical sense is one from
which it is costly to exclude users but whose units are subtractable, so that one
appropriator's use diminishes what remains for others; the governance problem is to
align appropriation and provision so that the resource is neither overused nor
under-maintained. The distinction is load-bearing for the present article because
it licenses treating the AI stack as a governance design problem rather than as a
tragedy to be averted only through enclosure or top-down regulation.

The second move is analytical. From her comparative field studies
\citet{ostrom1990governing} distilled a set of design principles that
long-enduring self-governed common-pool-resource institutions tend to share, and
that institutions which fail tend to violate. These principles are the columns of
our taxonomy, so we state them in the form in which the taxonomy uses them.

\begin{principle}[Clearly defined boundaries]
The boundaries of the resource and the set of individuals or groups entitled to
appropriate from it are clearly defined. For an AI commons this concerns both what
counts as the pooled resource, a corpus, a weight set, a compute allocation, and
who is a member of the governing community.
\end{principle}

\begin{principle}[Congruence between rules and local conditions]
Appropriation rules restricting time, place, technology, and quantity are
congruent with local conditions and with provision rules requiring contributions of
labor, money, or materials. A commons that lets members take without contributing,
or that imposes provision burdens unrelated to benefit, is unstable.
\end{principle}

\begin{principle}[Collective-choice arrangements]
Most individuals affected by the operational rules can participate in modifying
them. This is the principle that most sharply separates a commons from both a firm,
where rules are set by ownership, and a regulator, where they are set by statute.
\end{principle}

\begin{principle}[Monitoring]
Monitors who audit resource conditions and appropriator behavior are accountable to
the appropriators or are the appropriators themselves. In the AI setting,
monitoring maps onto transparency and auditability of data provenance, model
behavior, and compute and energy use.
\end{principle}

\begin{principle}[Graduated sanctions]
Appropriators who violate operational rules are subject to graduated sanctions,
proportionate to the seriousness and context of the offense, administered by other
appropriators or by accountable officials.
\end{principle}

\begin{principle}[Conflict-resolution mechanisms]
Appropriators and their officials have rapid access to low-cost local arenas for
resolving conflicts among themselves or with officials.
\end{principle}

\begin{principle}[Minimal recognition of rights to organize]
The right of appropriators to devise their own institutions is not challenged by
external governmental authorities. For AI commons this concerns the legal
recognition of trusts, cooperatives, foundations, and licenses as vehicles of
self-governance.
\end{principle}

\begin{principle}[Nested enterprises]
For commons that are parts of larger systems, appropriation, provision, monitoring,
enforcement, conflict resolution, and governance are organized in multiple nested
layers. This is the principle of polycentricity, and it is the one that scales a
local AI commons into an ecosystem.
\end{principle}

The analytic engine beneath these principles is the Institutional Analysis and
Development framework, which decomposes any governance situation into an action
arena conditioned by the biophysical attributes of the resource, the attributes of
the community, and the rules-in-use, and which resolves the rules into position,
boundary, choice, aggregation, information, payoff, and scope rules
\citep{ostrom2005understanding,ostrom2011background}. The framework is a diagnostic
scaffold rather than a predictive theory, and it supplies the dimensions along which
our taxonomy compares institutions, so that the axes are inherited from a validated
grammar rather than invented for the occasion.

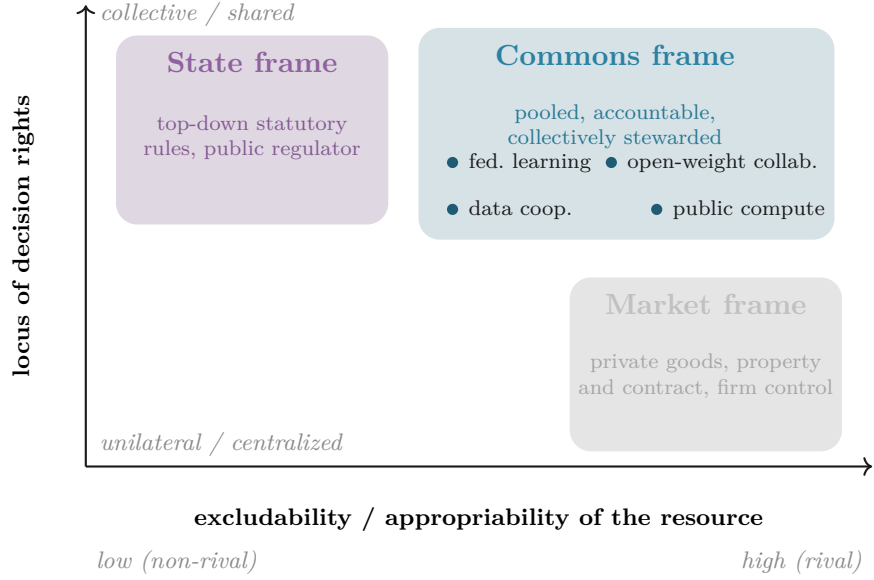
\begin{figure}[t]
\centering
\begin{tikzpicture}[scale=1.0, every node/.style={font=\small}]
  \draw[->,thick,pInk] (0,0) -- (10.4,0) node[below left=2pt and -6pt]{};
  \draw[->,thick,pInk] (0,0) -- (0,6.1);
  \node[rotate=90,anchor=south,font=\footnotesize\bfseries] at (-0.55,3.05)
        {locus of decision rights};
  \node[font=\footnotesize\bfseries] at (5.2,-0.7)
        {excludability / appropriability of the resource};
  \node[font=\footnotesize\itshape,pMute,anchor=west] at (0.05,5.98) {collective / shared};
  \node[font=\footnotesize\itshape,pMute,anchor=west] at (0.05,0.30) {unilateral / centralized};
  \node[font=\footnotesize\itshape,pMute,anchor=south west] at (0.0,-1.55) {low (non-rival)};
  \node[font=\footnotesize\itshape,pMute,anchor=south east] at (10.4,-1.55) {high (rival)};
  \begin{scope}[on background layer]
    \fill[pMarket!22,rounded corners=8pt] (6.4,0.2) rectangle (10.0,2.5);
    \fill[pState!22,rounded corners=8pt] (0.4,3.2) rectangle (4.0,5.7);
    \fill[pCommons!18,rounded corners=8pt] (4.4,3.0) rectangle (9.9,5.8);
  \end{scope}
  \node[pMarket!140!black,font=\bfseries] at (8.2,2.15) {Market frame};
  \node[pMarket!120!black,font=\scriptsize,align=center] at (8.2,1.2)
        {private goods, property\\ and contract, firm control};
  \node[pState!130!black,font=\bfseries] at (2.2,5.35) {State frame};
  \node[pState!120!black,font=\scriptsize,align=center] at (2.2,4.35)
        {top-down statutory\\ rules, public regulator};
  \node[pCommons!140!black,font=\bfseries] at (7.0,5.45) {Commons frame};
  \node[pCommons!130!black,font=\scriptsize,align=center] at (7.0,4.5)
        {pooled, accountable,\\ collectively stewarded};
  \foreach \x/\y/\lab in {4.85/3.40/data coop., 4.85/4.02/fed.\ learning,
        7.55/3.40/public compute, 6.95/4.02/open-weight collab.}{
    \fill[pCommons] (\x,\y) circle (2.1pt);
    \node[font=\scriptsize,pInk,anchor=west] at (\x+0.08,\y) {\lab};
  }
\end{tikzpicture}
\caption{The three institutional frames of AI governance. The market frame places
decision rights with private owners of excludable goods; the state frame places
them with a central regulator; the commons frame occupies the upper region in which
decision rights are held collectively. Commons-governed AI institutions
(illustrative points) span a range of resource excludability, from near-non-rival
knowledge to highly rival frontier compute, while sharing the collective locus of
decision rights. The vertical position, not the horizontal, is what distinguishes a
commons; rivalry shapes how hard the commons is to sustain, following
\citet{ostrom1990governing}.}
\label{fig:frames}
\end{figure}

The third move extends the commons from natural resources to knowledge.
\citet{hess2007understanding} argue that information and knowledge can be analyzed
as commons even though their content is non-rival in consumption, because the
infrastructures that store, curate, and transmit them, and the labor that maintains
provenance and quality, are subtractable and congestible. This is the conceptual
hinge that lets us treat a training corpus, a weight set, or a benchmark as a
genuine commons rather than as a mere public good: copying a dataset does not
diminish it, but curating, documenting, hosting, defending its licensing, and
sustaining its provenance are rivalrous activities that demand provision rules. The
Governing Knowledge Commons program of \citet{madison2010constructing} and
\citet{frischmann2014governing} sharpens this insight into a structured case-study
method, observing that cultural and intellectual commons are constructed rather than
naturally bounded and that their resources are frequently produced by the very
sharing arrangement under study. Our per-archetype characterization mirrors this
interrogation, asking of each commons-governed AI institution how its boundaries are
constructed, who its contributor community is, what its openness rules are, and what
outcomes it produces.

The fourth move concerns production. \citet{benkler2002coase} introduces
commons-based peer production, arguing through the economics of the firm that
radically decentralized, non-proprietary collaboration can outperform both firms
and markets for information production when the work is modular, its components are
of fine and heterogeneous granularity, and the cost of integrating contributions is
low; free and open-source software is the canonical case, and the argument is
developed into a general account of the networked information economy in
\citet{benkler2006wealth}. The empirical counterpart is the large-N study of
free/libre and open-source projects by \citet{schweik2012internet}, which
identifies the conditions under which open-source commons succeed rather than are
abandoned. This strand matters because many AI commons are not merely governed pools
but are produced by peer production, and the taxonomy must distinguish the
governance of a resource from the peer-production process that creates it. Against
this constructive literature stands the critical counter-narrative of enclosure.
\citet{boyle2003second} frames the expansion of intellectual property as a second
enclosure movement fencing off the intangible public domain, and \citet{lessig2006code}
makes the foundational argument that technical architecture is itself a regulator,
that code is law. Both are essential for an AI commons, because the governance of an
AI resource is enforced not only by licenses and norms but by architecture, through
release modalities, interface gating, and the granting or withholding of weight
access, so that the taxonomy must treat architectural constraint as a governance
instrument on a par with rules-in-use. The most recent strand makes the extension to
AI explicit: \citet{purtova2024data} survey the data-commons literature and assess
governance models against the common-pool-resource tradition, and
\citet{linaker2022sustaining} apply Ostrom's design principles directly to open-data
ecosystems. These works establish that applying commons theory to AI resources is a
live research move rather than a speculative one, and they mark the frontier the
present taxonomy aims to systematize.

A fifth move situates the commons tradition within the older intellectual lineage
from which it descends, for the Ostromian programme is best read as the empirical
formalization of a long current of non-statist and self-organizing political thought
rather than as its origin. The proposition that durable order can arise from the free
federation of autonomous units, without recourse to a sovereign centre, is the core
of the mutualist and federalist arguments of \citet{proudhon1979federation}, and the
nested, multi-layer scaling that Ostrom names polycentricity is the institutional
descendant of that federative principle. The claim that cooperation, and not
competition alone, is a primary factor in social organization was advanced by
\citet{kropotkin1902mutual}, and it supplies the motivational substrate that the
design principles presuppose but do not themselves explain, namely why appropriators
contribute provision when narrow self-interest would counsel free-riding. The
confederal and ecological strand of \citet{bookchin1982ecology} ties the
self-government of resources to a critique of hierarchy and to ecological limits,
prefiguring both the polycentric scaling of the eighth design principle and the
treatment of energy and sustainability as governance problems internal to the
commons rather than as external constraints. That such self-organization is not
utopian but already pervasive in ordinary social life is the observation of
\citet{ward1973anarchy}, which underwrites the empirical posture of this article,
that commons-governed artificial intelligence describes existing practice and is not
merely a proposal. The decisive complement to this constructive lineage is the
critique of the high-modernist state offered by \citet{scott1998seeing}, whose
account of how centrally legible schemes fail when they suppress the local practical
knowledge he terms m\^etis supplies the argument, absent from a purely taxonomic
treatment, for why the state frame alone is insufficient and why decision rights
distributed across a polycentric order can succeed where a single planning centre
cannot. We invoke this lineage as analytic genealogy and not as program, since its
value is that the design principles acquire a rationale, the federative ancestry of
polycentricity, the cooperative basis of provision, and the epistemic case against
central legibility, that an enumeration of principles alone leaves implicit. The
present section has read that lineage analytically, as the source of the principles;
the next reads it chronologically, tracing how the same idea travelled from
nineteenth-century federation through the knowledge commons to the AI stack, in order
to show that commons-governed artificial intelligence is the continuation of a long
argument rather than a recent coinage.

%% file: sec_genealogy.tex
\section{A Genealogy of the Commons Idea, from Federation to Artificial Intelligence}
\label{sec:genealogy}

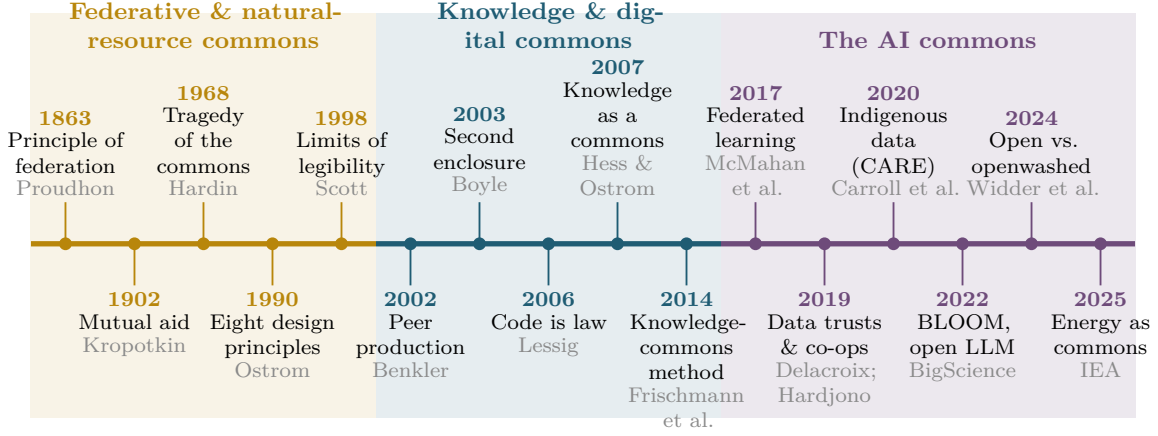
\begin{figure}[t]
\centering
\begin{tikzpicture}[font=\small]
  \definecolor{eA}{HTML}{B8860B}  
  \definecolor{eB}{HTML}{1F5C73}  
  \definecolor{eC}{HTML}{6E4B7A}  
  \def\W{14.6} \def\s{0.9125}     
  \begin{scope}[on background layer]
    \fill[eA!10] (0,-2.3)        rectangle (5*\s,3.05);
    \fill[eB!11] (5*\s,-2.3)     rectangle (10*\s,3.05);
    \fill[eC!13] (10*\s,-2.3)    rectangle (\W,3.05);
  \end{scope}
  \draw[eA,line width=2pt] (0,0)      -- (5*\s,0);
  \draw[eB,line width=2pt] (5*\s,0)   -- (10*\s,0);
  \draw[eC,line width=2pt] (10*\s,0)  -- (\W,0);
  \node[anchor=south,font=\footnotesize\bfseries,eA,align=center,text width=4.3cm]
        at (2.5*\s,2.45) {Federative \& natural-resource commons};
  \node[anchor=south,font=\footnotesize\bfseries,eB,align=center,text width=4.3cm]
        at (7.5*\s,2.45) {Knowledge \& digital commons};
  \node[anchor=south,font=\footnotesize\bfseries,eC,align=center,text width=4.3cm]
        at (13*\s,2.45) {The AI commons};
  \foreach \i/\yr/\er/\ds/\au in {%
      1/1863/eA/{Principle of federation}/Proudhon,
      3/1968/eA/{Tragedy of the commons}/Hardin,
      5/1998/eA/{Limits of legibility}/Scott,
      7/2003/eB/{Second enclosure}/Boyle,
      9/2007/eB/{Knowledge as a commons}/{Hess \& Ostrom},
      11/2017/eC/{Federated learning}/{McMahan et al.},
      13/2020/eC/{Indigenous data (CARE)}/{Carroll et al.},
      15/2024/eC/{Open vs.\ openwashed}/{Widder et al.}}{
    \pgfmathsetmacro{\x}{(\i-0.5)*\s}
    \draw[\er,line width=0.7pt] (\x,0) -- (\x,0.55);
    \fill[\er] (\x,0) circle (2.4pt);
    \node[anchor=south,text width=1.7cm,align=center,font=\scriptsize,inner sep=1pt]
         at (\x,0.6) {{\color{\er}\bfseries \yr}\\[-1pt] \ds\\[-1pt]
                      {\scriptsize\color{pMute}\au}};
  }
  \foreach \i/\yr/\er/\ds/\au in {%
      2/1902/eA/{Mutual aid}/Kropotkin,
      4/1990/eA/{Eight design principles}/Ostrom,
      6/2002/eB/{Peer production}/Benkler,
      8/2006/eB/{Code is law}/Lessig,
      10/2014/eB/{Knowledge-commons method}/{Frischmann et al.},
      12/2019/eC/{Data trusts \& co-ops}/{Delacroix; Hardjono},
      14/2022/eC/{BLOOM, open LLM}/BigScience,
      16/2025/eC/{Energy as commons}/IEA}{
    \pgfmathsetmacro{\x}{(\i-0.5)*\s}
    \draw[\er,line width=0.7pt] (\x,0) -- (\x,-0.55);
    \fill[\er] (\x,0) circle (2.4pt);
    \node[anchor=north,text width=1.7cm,align=center,font=\scriptsize,inner sep=1pt]
         at (\x,-0.6) {{\color{\er}\bfseries \yr}\\[-1pt] \ds\\[-1pt]
                      {\scriptsize\color{pMute}\au}};
  }
\end{tikzpicture}
\caption{A genealogy of the commons idea in three periods. The federative and
natural-resource period (gold) supplies the political theory of self-organization
and Ostrom's design principles; the knowledge and digital period (teal) migrates the
commons to non-rival information goods and builds the case-study apparatus; the AI
period (plum) applies that apparatus layer by layer to the AI stack. Events are
spaced evenly rather than to scale, so the year labels read the acceleration directly:
roughly a century separates the first two milestones, and one to two years separate
the most recent. The argument of this article systematizes the third period with the
grammar built in the first two.}
\label{fig:timeline}
\end{figure}

The previous section read the commons tradition analytically, as the source of the
design principles that form the columns of the taxonomy. It is worth reading the
same tradition a second time chronologically, because the question that motivates
this article, how a resource as novel as the AI stack could already have a mature
vocabulary of collective governance waiting for it, is answered by the long history
of the idea rather than by its logical structure. The vocabulary was not built for
artificial intelligence; it was built over more than a century and a half for
pastures, forests, software, and data, and it arrives at the AI stack as the latest
term in a sequence. Figure~\ref{fig:timeline} sets out that sequence in three
periods, and we trace it here to establish that commons-governed artificial
intelligence is the current chapter of a continuous argument and not a recent
coinage.

The first period runs from the middle of the nineteenth century to the end of the
twentieth and concerns natural resources and the political theory of
self-organization. The thesis that durable order can arise from the free federation
of autonomous units, without a sovereign centre, is stated as a principle of social
organization by \citet{proudhon1979federation}, whose argument that legitimate order
is built upward from federated units rather than imposed downward from a state is the
distant ancestor of what Ostrom would later call polycentricity. The claim that
cooperation is a factor of survival on a par with competition is advanced by
\citet{kropotkin1902mutual}, supplying the motivational premise, why members provide
to a commons rather than free-ride, that the design principles presuppose but do not
themselves prove. For most of the twentieth century these remained heterodox
positions, and the dominant analytic frame was instead the one crystallized by
\citet{hardin1968tragedy}, whose tragedy of the commons appeared to demonstrate that
shared resources must be either privatized or policed. The decisive empirical
correction is the comparative fieldwork of \citet{ostrom1990governing}, which
established that the resources Hardin described are in practice governed by bounded
communities under self-given rules, and which distilled the eight design principles;
the complementary epistemic argument, that centrally legible schemes fail when they
suppress local practical knowledge, is made by \citet{scott1998seeing}. By the close
of the century the proposition that an institutional space exists between the market
and the state was no longer a conjecture but a documented regularity.

The second period, roughly the first decade and a half of the present century, is the
migration of the commons from natural resources to information, and it is the period
that makes the application to artificial intelligence possible at all. The economic
argument that radically decentralized, non-proprietary collaboration can outproduce
both firms and markets for information goods is made by \citet{benkler2002coase} and
developed into a general account of the networked economy in \citet{benkler2006wealth}.
The counter-movement of enclosure is named in the same years: \citet{boyle2003second}
describes the expansion of intellectual property as a second enclosure of the
intangible public domain, and \citet{lessig2004free} documents how law and technology
together lock down culture, an argument he sharpens into the thesis that architecture
is itself a regulator, that code is law \citep{lessig2006code}. The conceptual hinge
on which this article turns is supplied in this period by \citet{hess2007understanding},
who argue that knowledge can be analyzed as a commons despite the non-rivalry of its
content, because the infrastructures and labor that curate and sustain it are
subtractable. The empirical study of free and open-source software by
\citet{schweik2012internet} and the case-study method of the Governing Knowledge
Commons program \citep{frischmann2014governing} close the period by turning the
knowledge-commons idea into a research instrument. By 2014 the commons had a fully
developed apparatus for non-rival, peer-produced, infrastructurally subtractable
resources, which is precisely what the layers of the AI stack would turn out to be.

The third period is the present one, in which the apparatus is applied to artificial
intelligence, and its defining feature is acceleration: the milestones that were
decades apart in the first period fall a year or two apart in the third, as
Figure~\ref{fig:timeline} makes visible. The stewardship principles for research data
are codified as FAIR \citep{wilkinson2016fair} and, for communities rather than
individuals, as Indigenous data sovereignty \citep{kukutai2016indigenous}. Federated
learning \citep{mcmahan2017communication} supplies the technical substrate for a
commons of learning where a commons of raw data is foreclosed. The data layer acquires
explicit institutional forms in the bottom-up data trust \citep{delacroix2019bottomup}
and the data cooperative \citep{hardjono2019datacoops}, and a normative content in the
CARE principles \citep{carroll2020care}. The model layer acquires its commons exemplar
in the openly built BLOOM \citep{bigscience2022bloom}, its accountability machinery in
the Foundation Model Transparency Index \citep{bommasani2023foundation}, and its
critical self-correction in the openwashing diagnosis of \citet{widder2024open} and the
Open Source AI Definition \citep{osi2024osaid}. The compute layer is articulated as
public infrastructure in the national and supranational initiatives
\citep{nairr2023strengthening,eurohpc2024aifactories}, and the energy layer becomes a
governance object at macro scale with the demand projections of the
\citet{iea2025energyai}. Within roughly a decade every layer of the stack acquired a
commons articulation.

What the chronology shows that the analytic reconstruction cannot is the direction and
the speed of transmission. The idea moved from pastures to bitstreams to model weights,
each migration carrying the same core, a bounded community giving itself rules over a
subtractable resource, into a domain its originators did not anticipate, and the
migrations grew closer together as the digital substrate spread. The AI-era instances
of the third period arrived, however, without the comparative grammar that the first
two periods had developed, each named in the vocabulary of its own subfield and studied
in isolation. The contribution of the present article is to supply that grammar, by
recognizing the third-period instances as the continuation of the first two and by
classifying them with the apparatus those periods produced. With the lineage of the
idea established, we turn to the object it is now being asked to govern, the layered
AI stack.

%% file: sec_stack.tex
\section{The AI Stack as a Layered Commons}
\label{sec:stack}

To classify how AI is governed in common we must first fix what is being governed.
The contemporary AI supply chain is conventionally decomposed into three core
layers, compute, data, and models, where compute denotes the physical and software
infrastructure of accelerators, interconnect, and data centers, data denotes the
corpora on which models are trained and evaluated, and models denotes the trained
artifacts and the weights that parameterize them. We retain this decomposition but
extend it in two directions that the commons frame makes necessary. Downward, we add
\emph{energy} as the physical substrate on which compute runs, because at current
scale the electricity and water drawn by AI training and inference have become
shared-resource conflicts in their own right \citep{iea2025energyai,masanet2020recalibrating}.
Laterally, we separate a \emph{knowledge and evaluation} layer from the model layer,
comprising the benchmarks, leaderboards, documentation standards, and open
toolchains that are produced by peer production and that determine what counts as
progress. The result is a five-layer view of the AI stack as a stratified commons,
each layer of which we now define and argue is genuinely commons-like.

\begin{definition}[Commons-governed AI institution]
A \emph{commons-governed AI institution} is an arrangement in which one or more
resource layers of the AI stack are pooled and stewarded as a shared resource whose
access, contribution, and benefit are determined by a defined community under
self-given rules, rather than by unilateral private control or top-down statutory
command. Equivalently, in the terms of Section~\ref{sec:commons}, it is an
institution whose decision rights over an AI resource are held collectively and
whose operation can be characterized by Ostrom's design principles.
\end{definition}

The qualifier that does the work in this definition is subtractability. A resource
need not be physically rival to be a commons; it need only be subtractable in some
governance-relevant dimension, so that uncontrolled appropriation or insufficient
provision degrades it \citep{hess2007understanding}. We argue layer by layer that
each of the five satisfies this criterion.

The \emph{data} layer comprises the training and evaluation corpora. Their
informational content is non-rival in copying, but the activities that make a corpus
usable, curation, deduplication, documentation, consent management, license defense,
and the maintenance of provenance, are rivalrous in labor and congestible in
infrastructure, and the supply of high-quality openly licensed data is in fact
shrinking under enclosure pressure as rights-holders restrict reuse for AI training
\citep{purtova2024data}. A corpus left ungoverned degrades through quality
deterioration, contamination, and the erosion of consent, which is exactly the
provision-and-appropriation problem Ostrom's principles address.

The \emph{compute} layer comprises accelerators, clusters, and the cloud
infrastructure that schedules them. Compute is the most clearly rival of the five
layers and, as \citet{sastry2024computing} argue, the most governable node of the
supply chain, because it is detectable, excludable, quantifiable, and produced
through an extremely concentrated manufacturing pipeline. These same properties make
compute both the most natural candidate for enclosure and the layer where a commons
is hardest to sustain: a resource that is easy to meter and to deny is a resource
around which exclusion is the default. The access asymmetry this produces, the
compute divide between the few organizations that command frontier-scale clusters
and everyone else, is the inequality that public-compute and compute-commons
proposals exist to redress \citep{heim2024governing,nairr2023strengthening}.

The \emph{model} layer comprises trained weights and the interfaces through which
they are served. Weights are non-rival once released, but their governance is
subtractable through the architectural channel identified by
\citet{lessig2006code}: a developer chooses a release modality, from a closed
inference API through gated weights to fully open weights, and that choice is a
governance act that the community around the model may or may not control. The
transparency of the upstream resources that produced a model, its data, its data
labor, and its computational cost, is itself a scarce and contested good, as the low
scores on exactly those indicators in the Foundation Model Transparency Index
demonstrate \citep{bommasani2023foundation,bommasani2024foundation}.

The \emph{knowledge and evaluation} layer comprises benchmarks, evaluation suites,
documentation standards such as model cards \citep{mitchell2019model}, and the open
software toolchains on which the field runs. This is the layer most fully explained
by commons-based peer production \citep{benkler2002coase}: benchmarks and libraries
are produced by loosely coordinated contributors, are non-rival in use, and are
nonetheless subtractable in maintenance, because an unmaintained benchmark becomes
contaminated and an unmaintained library bit-rots. The governance question is who
sets the standard of progress and who bears the upkeep.

The \emph{energy} layer comprises the electricity, and relatedly the water for
cooling, consumed by training and inference. Energy is unambiguously rival and its
appropriation is a classical commons problem of siting, grid impact, and carbon
intensity. Per-model accounting has made the cost legible, from the training-time
emissions measured by \citet{strubell2019energy} and \citet{patterson2021carbon} to
the full life-cycle assessment of \citet{luccioni2023estimating}, and the macro-scale
demand projections of \citet{iea2025energyai} turn that cost into a governance
problem at the level of national grids. Treating energy as a layer of the AI commons
rather than as an external cost is one of the article's organizing choices, and it
connects AI governance to the established community-energy commons tradition
\citep{wade2025energy}.

\begin{figure}[t]
\centering
\begin{tikzpicture}[font=\small]
  \def\w{11.4}
  \definecolor{Lk}{HTML}{1F5C73}
  \definecolor{Lm}{HTML}{3E7C8C}
  \definecolor{Ld}{HTML}{6FA8B8}
  \definecolor{Lc}{HTML}{B8860B}
  \definecolor{Le}{HTML}{8C6508}
  \foreach \i/\col/\name/\desc in {
      0/Le/ENERGY/{electricity \& water; grid impact, carbon intensity},
      1/Lc/COMPUTE/{accelerators, clusters, cloud scheduling},
      2/Ld/DATA/{training \& evaluation corpora, provenance, consent},
      3/Lm/MODELS/{weights, release modality, serving interface},
      4/Lk/{KNOWLEDGE \& EVAL.}/{benchmarks, toolchains, documentation}}{
    \fill[\col!85] (0,\i*1.06) rectangle (\w,\i*1.06+0.96);
    \node[white,font=\bfseries,anchor=west] at (0.25,\i*1.06+0.66) {\name};
    \node[white,font=\scriptsize,anchor=west] at (0.25,\i*1.06+0.30) {\desc};
  }
  \draw[-{Latex[length=3mm]},very thick,pState] (\w+0.35,0.1) -- (\w+0.35,5.2);
  \node[rotate=90,font=\scriptsize\bfseries,pState!130!black] at (\w+0.92,2.65)
      {enclosure pressure};
  \draw[-{Latex[length=3mm]},very thick,pCommons] (-0.35,5.2) -- (-0.35,0.1);
  \node[rotate=90,font=\scriptsize\bfseries,pCommons!130!black] at (-0.92,2.65)
      {commons pooling};
  \node[anchor=north west,font=\scriptsize\itshape,pMute,text width=\w cm,align=left] at (0.0,-0.45)
     {Each layer is subtractable in a governance-relevant dimension even where its informational content is non-rival.};
\end{tikzpicture}
\caption{The AI stack as a five-layer commons. Energy is the physical substrate;
compute, data, and models are the conventional supply-chain layers; knowledge and
evaluation is the peer-produced layer that fixes the standard of progress. The
default pressure on every layer is enclosure (right), against which commons
arrangements pool and collectively steward the resource (left). The taxonomy of
Section~\ref{sec:taxonomy} crosses these five layers with the eight design
principles of Section~\ref{sec:commons}.}
\label{fig:stack}
\end{figure}

These five layers are not independent. Energy constrains compute, compute and data
jointly produce models, and the knowledge layer determines how all of the others are
measured and improved. A consequence we develop in Section~\ref{sec:synthesis} is
that commons governance applied to a single layer in isolation is fragile: open
weights served from enclosed compute trained on opaque data leave the underlying
concentration of power intact, which is precisely the critique that
\citet{widder2024open} level at nominal openness. The layered view therefore does
double duty, fixing the first axis of the taxonomy and motivating its central
empirical claim that durable AI commons are cross-layer rather than single-layer.

%% file: sec_taxonomy.tex
\section{Constructing the Taxonomy}
\label{sec:taxonomy}

A taxonomy is useful to the degree that its dimensions are conceptually grounded,
mutually exclusive enough to classify, and collectively exhaustive enough to cover
the phenomenon. We build ours on two primary dimensions, both inherited from the
commons tradition rather than constructed for the occasion, and we record for each
institution a set of secondary descriptive attributes that capture variation the two
primary axes do not. The construction follows the iterative logic of established
taxonomy-development method \citep{nickerson2013method}, alternating between the
conceptual derivation of dimensions from the Ostromian grammar and the empirical
inspection of the institutions examined in Sections~\ref{sec:data}--\ref{sec:energy},
and it adopts the interrogation structure of the Governing Knowledge Commons
case-study method \citep{frischmann2014governing}, which asks of every commons how
its resource and community boundaries are constructed, what its rules-in-use are, and
what outcomes it produces.

The first primary dimension is the \emph{resource layer} of the AI stack that an
institution pools, taking values in the five-element set fixed in
Section~\ref{sec:stack}: data, compute, models, knowledge and evaluation, and energy.
A given institution may pool more than one layer, and indeed Section~\ref{sec:synthesis}
argues that the durable institutions are precisely those that pool several; the layer
dimension is therefore polythetic, recording the subset of layers an institution
governs rather than assigning it to a single class.

\begin{figure}[t]
\centering
\begin{tikzpicture}[font=\scriptsize]
  \def\dx{1.30} \def\dy{0.80}
  \def\funcs{{"B","A","C","M","S","R","L","N"}}
  \def\layers{{"ENERGY","KNOW/EVAL","MODELS","COMPUTE","DATA"}}
  \foreach \j/\lab in {0/B,1/A,2/C,3/M,4/S,5/R,6/L,7/N}{
     \node[font=\bfseries,pTeal] at (\j*\dx+\dx/2, 5*\dy+0.35) {\lab};
  }
  \foreach \i/\lab in {0/ENERGY,1/{KNOW \& EVAL},2/MODELS,3/COMPUTE,4/DATA}{
     \node[anchor=east,font=\bfseries,pInk] at (-0.12,\i*\dy+\dy/2) {\lab};
  }
  \foreach \i in {0,...,4}{
    \foreach \j in {0,...,7}{
      \draw[pMute!50,fill=pBg] (\j*\dx,\i*\dy) rectangle (\j*\dx+\dx,\i*\dy+\dy);
    }
  }
  \foreach \j in {0,1,2,3,6}{
     \fill[pGold!55] (\j*\dx+0.06,4*\dy+0.06) rectangle (\j*\dx+\dx-0.06,4*\dy+\dy-0.06);
  }
  \foreach \j in {0,1,3,6,7}{
     \fill[pTeal!45] (\j*\dx+0.06,3*\dy+0.06) rectangle (\j*\dx+\dx-0.06,3*\dy+\dy-0.06);
  }
  \foreach \j in {0,2,3,7}{
     \fill[pTealL!60] (\j*\dx+0.06,2*\dy+0.06) rectangle (\j*\dx+\dx-0.06,2*\dy+\dy-0.06);
  }
  \node[anchor=north west,font=\scriptsize,text width=10.4cm,align=left] at (0,-0.62)
     {\tikz\fill[pGold!55](0,0)rectangle(0.30,0.22);\ data cooperative \quad
      \tikz\fill[pTeal!45](0,0)rectangle(0.30,0.22);\ public compute commons \\[2pt]
      \tikz\fill[pTealL!60](0,0)rectangle(0.30,0.22);\ open-weight collaboration};
  \node[anchor=north west,font=\scriptsize\itshape,pMute,text width=10.4cm,align=left] at (0,-1.62)
     {Functions: B boundaries, A appropriation/provision, C collective choice, M monitoring,
      S sanctions, R conflict resolution, L recognition, N nesting.};
\end{tikzpicture}
\caption{The classificatory space of commons-governed AI. Rows are the five resource
layers of the AI stack (Figure~\ref{fig:stack}); columns are the eight Ostromian
governance functions (Section~\ref{sec:commons}). An institution occupies a region
of the grid rather than a single cell: three archetypes are overlaid to show that the
data cooperative concentrates in the data row, the public compute commons in the
compute row, and the open-weight collaboration in the model row, each implementing a
characteristic subset of the eight functions. The maturity matrix of
Figure~\ref{fig:heatmap} aggregates the field's coverage of every cell.}
\label{fig:grid}
\end{figure}
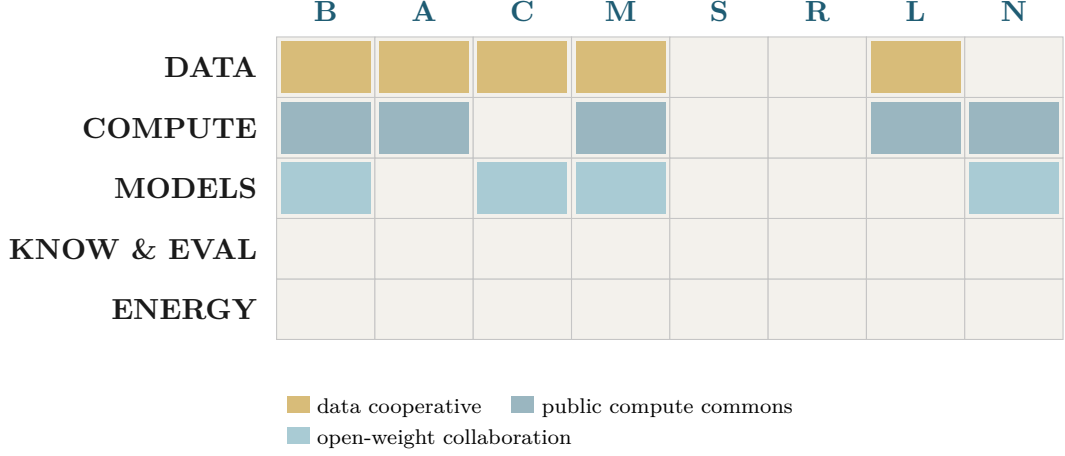

The second primary dimension is the \emph{governance function} an institution
performs, taking values in the eight design principles of
Section~\ref{sec:commons}: boundary definition, congruence of appropriation and
provision, collective choice, monitoring, graduated sanctions, conflict resolution,
recognition of the right to organize, and nested polycentric scaling. For each
(layer, function) pair an institution either does or does not implement the function
over the layer, and the strength with which it does so is what the maturity matrix of
Section~\ref{sec:synthesis} records. Crossing the five layers with the eight
functions yields the forty-cell classificatory space of Figure~\ref{fig:grid}, in
which any commons-governed AI institution occupies a region rather than a point: a
federated health consortium, for example, is concentrated in the data and compute
rows and is strong on boundaries, monitoring, and collective choice but weak on
graduated sanctions, while a public compute initiative occupies the compute row and
is strong on monitoring and recognition but weak on collective choice.

Two primary axes do not exhaust the variation among commons-governed AI
institutions, so we record four secondary descriptive attributes that the layer
sections fill in and that the synthesis collects in Table~\ref{tab:archetypes}. The
first is the position on the \emph{openness spectrum}, which ranges from a closed
proprietary resource, through a club good available to members on terms, through
genuinely open access, to a fully commons-governed resource that is both open and
collectively stewarded; the distinction between mere open access and a governed
commons, drawn in Section~\ref{sec:commons}, is exactly the distinction the upper end
of this spectrum encodes, and Section~\ref{sec:models} shows that openness without
collective governance is the failure mode \citet{widder2024open} diagnose. The second
is the \emph{legal vehicle} through which the institution holds rights and bears
duties, ranging across the trust, the cooperative, the foundation or non-profit
association, the public agency or intergovernmental body, the contractual license,
and the decentralized autonomous organization \citep{hassan2021dao}; the legal
vehicle is the concrete form Ostrom's seventh principle, recognition of the right to
organize, takes in practice. The third is the rule for \emph{benefit distribution},
which determines whether value created by the commons flows to contributors, to all
members equally, to the broader public, or to a steward, and which distinguishes a
cooperative that returns surplus to members from a public initiative that returns it
to society at large. The fourth is the \emph{sustainability stance}, recording
whether and how an institution accounts for and governs the energy and carbon cost of
its operation, an attribute we elevate to first-class status because
Section~\ref{sec:energy} argues that monitoring of energy use is a provision
obligation of any compute-bearing commons and not an optional add-on.

The classification produced by these dimensions is polythetic and morphological
rather than monothetic: an institution is characterized by the profile of values it
takes across layers, functions, and secondary attributes, not assigned to one of a
small number of mutually exclusive boxes. This is a deliberate choice forced by the
phenomenon, because the institutions examined below recombine the same governance
functions over different layers, and a monothetic scheme would either multiply
classes uncontrollably or suppress the cross-layer combination that is the empirical
heart of the subject. The ten archetypes named in Section~\ref{sec:synthesis} are not
the classes of the taxonomy but recurrent regions of the space, useful as landmarks
precisely because real institutions cluster near them while varying in the secondary
attributes.

As a reading map for the evidence that follows, Table~\ref{tab:litmap} groups the
literature the taxonomy draws on by the region of the classificatory space each body
of work informs, separating the conceptual foundations and the AI-governance context
that supply the two axes from the five resource layers that the next sections examine
in turn. The grouping is itself an application of the taxonomy, in that a work is
placed by the layer and function it speaks to rather than by its disciplinary origin,
and the recurrence of several works across cells is the bibliographic trace of the
cross-layer character that Section~\ref{sec:synthesis} argues is constitutive of the
durable commons.

\begin{table}[tp]
\centering
\small
\setlength{\tabcolsep}{5pt}
\renewcommand{\arraystretch}{1.15}
\caption{A map of the literature by region of the taxonomy. Each row lists the works
that inform a given conceptual area or resource layer; works that bear on more than one
region recur. The five resource-layer blocks correspond to the rows of
Figures~\ref{fig:stack} and~\ref{fig:grid}, and the two conceptual blocks supply,
respectively, the function axis (the commons tradition) and the comparative backdrop
(the AI-governance context) against which the taxonomy is positioned.}
\label{tab:litmap}
\begin{tabularx}{\textwidth}{@{}>{\raggedright\arraybackslash}p{4.4cm} X@{}}
\toprule
\textbf{Thematic cluster} & \textbf{Contributing literature} \\
\midrule
\multicolumn{2}{@{}l}{\textbf{Conceptual foundations: the commons tradition}}\\
\addlinespace[2pt]
Common-pool resources, polycentricity & \citet{hardin1968tragedy}; \citet{ostrom1990governing,ostrom2005understanding,ostrom2010beyond,ostrom2011background} \\
Knowledge as a commons & \citet{hess2007understanding}; \citet{madison2010constructing}; \citet{frischmann2014governing} \\
Peer production and enclosure & \citet{benkler2002coase,benkler2006wealth}; \citet{boyle2003second}; \citet{lessig2004free,lessig2006code}; \citet{schweik2012internet} \\
Federative, self-organization lineage & \citet{proudhon1979federation}; \citet{kropotkin1902mutual}; \citet{ward1973anarchy}; \citet{bookchin1982ecology}; \citet{scott1998seeing}; \citet{kleiner2010telekommunist} \\
Commons applied to data and AI & \citet{purtova2024data}; \citet{linaker2022sustaining} \\
\addlinespace[3pt]
\multicolumn{2}{@{}l}{\textbf{AI-governance context}}\\
\addlinespace[2pt]
Principles, ethics, risk & \citet{jobin2019global}; \citet{floridi2018ai4people}; \citet{mittelstadt2019principles}; \citet{slattery2024airisk} \\
State and regulatory frames & \citet{euaiact2024}; \citet{nist2023airmf}; \citet{oecd2019recommendation} \\
\addlinespace[3pt]
\multicolumn{2}{@{}l}{\textbf{Resource layer: data}}\\
\addlinespace[2pt]
Trusts, cooperatives, sovereignty & \citet{delacroix2019bottomup}; \citet{hardjono2019datacoops}; \citet{micheli2020emerging}; \citet{carroll2020care}; \citet{kukutai2016indigenous} \\
Open corpora, standards, practice & \citet{ardila2020commonvoice}; \citet{schuhmann2022laion5b}; \citet{wilkinson2016fair}; \citet{chafetz2024blueprint}; \citet{openfuture2024commons} \\
\addlinespace[3pt]
\multicolumn{2}{@{}l}{\textbf{Resource layer: compute}}\\
\addlinespace[2pt]
Compute governance, public pools & \citet{sastry2024computing}; \citet{heim2024governing}; \citet{nairr2023strengthening}; \citet{eurohpc2024aifactories} \\
Federated learning & \citet{mcmahan2017communication}; \citet{kairouz2021advances} \\
\addlinespace[3pt]
\multicolumn{2}{@{}l}{\textbf{Resource layer: models}}\\
\addlinespace[2pt]
Foundation models and openness & \citet{bommasani2021opportunities,bommasani2023foundation,bommasani2024foundation}; \citet{osi2024osaid}; \citet{widder2024open}; \citet{kapoor2024societal} \\
Collaboration, documentation, vehicles & \citet{bigscience2022bloom}; \citet{mitchell2019model}; \citet{hassan2021dao} \\
\addlinespace[3pt]
\multicolumn{2}{@{}l}{\textbf{Resource layer: knowledge and evaluation}}\\
\addlinespace[2pt]
Peer-produced standards, benchmarks & \citet{benkler2002coase}; \citet{schweik2012internet}; \citet{mitchell2019model}; \citet{wilkinson2016fair} \\
\addlinespace[3pt]
\multicolumn{2}{@{}l}{\textbf{Resource layer: energy}}\\
\addlinespace[2pt]
Footprint measurement & \citet{strubell2019energy}; \citet{patterson2021carbon}; \citet{luccioni2023estimating}; \citet{masanet2020recalibrating} \\
Macro demand, community energy & \citet{iea2025energyai}; \citet{wade2025energy}; \citet{radovanovic2023carbon}; \citet{schwartz2020green}; \citet{henderson2020towards} \\
\addlinespace[3pt]
\multicolumn{2}{@{}l}{\textbf{Method and public-interest framing}}\\
\addlinespace[2pt]
Taxonomy development & \citet{nickerson2013method} \\
Public-interest AI & \citet{publicai2024whitepaper} \\
\bottomrule
\end{tabularx}
\end{table}

With the axes fixed, we turn to the evidence, examining each resource
layer in the order data, compute, models, knowledge, and energy.

%% file: sec_data.tex
\section{The Data Layer}
\label{sec:data}

The data layer is the most institutionally developed of the five, because the
governance of data predates the foundation-model era and arrives in the AI commons
with a mature vocabulary of trusts, cooperatives, and sovereignty regimes. The
organizing question of the layer is who holds the rights to pool, curate, and license
a corpus, and to whose benefit the resulting models flow. Four families of
arrangement answer it in commons terms, and each instantiates a recognizable subset
of Ostrom's design principles over the data row of the taxonomy.

The data trust is a fiduciary arrangement in which data subjects pool rights of
control over their data and vest them in a trustee who is legally bound to exercise
those rights in the subjects' interest. \citet{delacroix2019bottomup} develop the
bottom-up data trust as a deliberate alternative to a one-size-fits-all default,
arguing for an ecology of trusts with differing terms among which subjects can choose
and to which they can defect, so that the trust structure supplies the
collective-choice and conflict-resolution functions that bilateral consent between an
individual and a platform cannot. In taxonomic terms the trust is strong on
recognition of the right to organize, because the trust is a well-understood legal
vehicle, and on boundary definition, because membership and the pooled resource are
defined by the trust instrument, while its collective-choice strength depends on
whether beneficiaries can instruct the trustee or merely exit. The trust answers the
fiduciary-duty objection that \citet{mittelstadt2019principles} raises against
principlism, supplying exactly the enforceable duty that a list of principles lacks.

The data cooperative places curation and analytics under member ownership.
\citet{hardjono2019datacoops} model it on the credit union, a member-owned fiduciary
that aggregates members' personal data, runs analytics on their behalf, and returns
the resulting value to the membership, so that the cooperative is strong on
collective choice and on benefit distribution to contributors, the two functions
that most sharply distinguish it from a firm. The cooperative form realizes Ostrom's
third principle directly, because members participate in modifying the operational
rules through democratic governance, and it realizes the second principle, congruence
of appropriation and provision, because members both contribute data and share in the
surplus. Its characteristic weakness is on graduated sanctions and monitoring at
scale, since a large cooperative struggles to audit member behavior, and on the
provision of the compute needed to act on the pooled data, which is why
Section~\ref{sec:synthesis} reads the cooperative as a data-row institution that
becomes durable only when coupled to a compute-row arrangement.

The closest prior taxonomy of the layer is that of \citet{micheli2020emerging}, who
identify four emerging models of data governance, data-sharing pools, data
cooperatives, public data trusts, and personal data sovereignty, classified by who
holds power and to whose benefit data flows. That taxonomy is the proximate ancestor
of the present one and the reason we position carefully against it: it is organized by
actor and power rather than by commons-governance mechanism, it predates the
foundation-model era, and it treats data in general rather than AI training assets in
particular. Our contribution at this layer is to re-read its four models as profiles
over the eight Ostromian functions and to extend the analysis from personal data to
the training corpora and provenance infrastructures that foundation models consume.
The survey of \citet{purtova2024data} supplies the bridge, distinguishing data as an
economic good from data as a commons and assessing the governance models against the
common-pool-resource tradition, and it documents the enclosure pressure, the
shrinking supply of openly licensed data, that makes a data commons urgent rather than
optional.

The cultural and community strand is where the data layer most clearly exceeds
individual consent. The CARE principles for Indigenous data governance,
\emph{collective benefit}, \emph{authority to control}, \emph{responsibility}, and
\emph{ethics}, articulated by \citet{carroll2020care}, assert that data about a people
are governed by that people as a collective, not merely by the individuals it
describes, and they were formulated as a complement to the stewardship-oriented FAIR
principles of \citet{wilkinson2016fair}, which make data findable, accessible,
interoperable, and reusable but say nothing about who decides. Indigenous data
sovereignty \citep{kukutai2016indigenous} is the broader movement of which CARE is the
principled expression, and it grounds the taxonomy's claim that commons governance is
not reducible to open access: a corpus can be fully open and still violate collective
authority, and a commons can legitimately restrict appropriation to honor it. This
strand is strong on collective choice and on boundary definition in the communal
sense, and it supplies the normative content that distinguishes a commons from a
public good.

The empirical referents of the layer are the large open corpora on which the claim
that AI training data can be commons-governed must ultimately be tested. The Common
Voice project of \citet{ardila2020commonvoice} assembles a massively multilingual
speech corpus through voluntary contribution and a permissive dedication to the public
domain, and it is among the cleanest existing instances of commons-based peer
production applied to AI training data, strong on provision through contribution and
on open boundaries. The LAION-5B image-text dataset of \citet{schuhmann2022laion5b},
by contrast, illustrates the governance questions a large open corpus raises rather
than settles, since its openness coexists with contested questions of consent,
provenance, and harmful content that a genuine data commons would have to govern
through monitoring and graduated sanctions it did not originally possess. Read
together, the two cases mark the range of the data layer: openness is necessary for a
data commons but not sufficient, and the difference between Common Voice and the
controversies around large scraped corpora is precisely the difference between a
governed commons and mere open access that Section~\ref{sec:commons} insists upon. The
practitioner literature, including the GovLab blueprint for unlocking new data commons
for AI \citep{chafetz2024blueprint} and the Open Future proposal for commons-based
dataset governance \citep{openfuture2024commons}, confirms that the category is being
actively built, while remaining, as practitioner reports, without the comparative
institutional structure this taxonomy supplies.

%% file: sec_compute.tex
\section{The Compute Layer}
\label{sec:compute}

Compute is the layer at which the gap between the promise of an AI commons and the
reality of concentration is widest. \citet{sastry2024computing} argue that computing
power is a uniquely governable node of the AI supply chain because it is detectable,
excludable, quantifiable, and produced through a manufacturing pipeline of extreme
concentration, in which a handful of firms control design, fabrication, and the
critical equipment upstream of them. The argument cuts two ways for the commons
project. The properties that make compute a lever for state and corporate control,
excludability and concentration, are the same properties that make a compute commons
both necessary and difficult, because a resource that is trivial to meter and to deny
is a resource around which enclosure is the default equilibrium and pooling is the
exception that must be deliberately constructed and defended. \citet{heim2024governing}
sharpen the institutional picture by locating cloud compute providers as regulatory
intermediaries, with latent capacities as securers, record keepers, verifiers, and
enforcers, and we read those capacities as the plumbing through which either enclosure
or commons-style stewardship must operate: the same provider that can enforce a state
export control can, under different governance, supply the monitoring and graduated
sanctions a compute commons requires.

The asymmetry these analyses describe is the compute divide, the concentration of
frontier-scale training capacity in a few organizations while public-interest users,
academic researchers, smaller firms, and the global majority operate at orders of
magnitude less. The divide is not merely a matter of cost but of governance, because
the organizations that command the compute also command the agenda of what is built
and evaluated, and it is the inequality that the public-compute proposals of the layer
exist to redress. The United States National AI Research Resource is the most
developed national articulation \citep{nairr2023strengthening}, proposing a
cooperative-stewardship model in which a federal mix of computational and data
resources is made accessible through an integrated portal with the explicit aims of
spurring innovation, diversifying the talent base, improving capacity, and advancing
trustworthy AI. In taxonomic terms the Resource is a state-anchored quasi-commons:
it pools provision and exercises collective choice through an advisory governance
structure, and it is strong on monitoring and on recognition of the right to organize,
but it stops short of the user self-governance that Ostrom's third principle places at
the center, since eligibility and allocation are set administratively rather than by
the user community. That gap is exactly the distinction the taxonomy uses to separate
a public-provision cell from a fully commons-governed one, and it is a productive gap
rather than a defect, because it locates the Resource precisely on the map.

The European counterpart supplies a second institutional form. The EuroHPC Joint
Undertaking, which pools publicly owned supercomputing capacity across member states,
was extended by a 2024 amendment to establish AI Factories that repurpose that
capacity for AI training and offer it to European users \citep{eurohpc2024aifactories}.
The Undertaking is federated across national and Union levels, which makes it the
clearest existing instance of Ostrom's eighth principle, nested enterprise, applied to
compute: appropriation and provision are organized in layers, with national centers
nested within a Union-level coordinating body. Together the United States and European
initiatives establish that large-scale public compute provision is no longer
hypothetical, which lets the taxonomy treat the compute commons as an empirical
category rather than a thought experiment, while their shared distance from user
self-governance marks the frontier the agenda of Section~\ref{sec:agenda} addresses.

A distinct and technically grounded route to a compute commons does not pool
hardware at all but pools learning. Federated learning, introduced by
\citet{mcmahan2017communication} as communication-efficient training of a shared model
over decentralized data through iterative averaging of locally computed updates, and
surveyed comprehensively by \citet{kairouz2021advances}, lets a community train a
common model without centralizing either the raw data or, in principle, the compute,
since each participant contributes the computation local to its own data. Federated
learning is therefore the substrate that makes a commons of \emph{learning} feasible
where a commons of raw data is blocked by privacy, regulation, or sovereignty, and it
recurs in the sectoral applications of Section~\ref{sec:applications}, most
prominently in health, where data cannot leave the institution that holds it. In
taxonomic terms a federated consortium is strong on boundary definition and on
monitoring, because participation and contribution are explicit and auditable, and its
collective-choice strength depends on whether the federation's governance lets
participants set the training objective and the aggregation rule or merely opt in to
one set by a coordinator. Its characteristic weaknesses are graduated sanctions, since
excluding a misbehaving participant is the only available sanction and is rarely
graduated, and the governance of the aggregate model's weights, which returns us to
the model layer. The compute layer thus exhibits in concentrated form the article's
central claim that single-layer commons are fragile: pooled compute that trains a
model whose weights are then enclosed has redistributed effort without redistributing
power, and a durable compute commons must reach across to the model and data layers it
serves.

%% file: sec_models.tex
\section{The Model Layer}
\label{sec:models}

The model layer governs trained weights and the interfaces through which they are
served, and it is the layer on which the meaning of openness is most contested. The
foundation-model paradigm named by \citet{bommasani2021opportunities} concentrated
capability into a small number of large artifacts whose governance is determined by a
single developer choice, the release modality, which ranges along a spectrum from a
fully closed inference interface to fully open weights and beyond. Because that choice
is an exercise of the architectural power that \citet{lessig2006code} identified as a
form of law, the model layer is where the taxonomy's secondary openness dimension does
the most work, and where the distinction between mere open access and a governed
commons becomes a practical design problem rather than a conceptual one.

\begin{figure}[t]
\centering
\begin{tikzpicture}[font=\small]
  \def\w{14.0}
  \shade[left color=pMute!55, right color=pCommons!75] (0,0) rectangle (\w,0.62);
  \draw[pInk] (0,0) rectangle (\w,0.62);
  \node[anchor=west,font=\scriptsize\itshape,white] at (0.18,0.31) {enclosure};
  \node[anchor=east,font=\scriptsize\itshape,white] at (\w-0.18,0.31) {collective stewardship};
  \foreach \x/\lab in {1.0/{Closed\\ API},4.2/{Gated\\ weights},7.4/{Open\\ weights},
        10.4/{Open-source AI\\ \scriptsize(OSI v1.0)},13.1/{Commons-\\ governed}}{
     \draw[pInk] (\x,0) -- (\x,-0.18);
     \node[anchor=north,align=center,font=\scriptsize] at (\x,-0.22) {\lab};
  }
  \node[anchor=north,align=center,font=\scriptsize,pMute,text width=3.0cm]
        at (1.0,-1.05) {governance internal to the developer};
  \node[anchor=north,align=center,font=\scriptsize,pCommons!130!black,text width=3.2cm]
        at (13.1,-1.05) {open \emph{and} collectively governed across layers};
  \draw[decorate,decoration={brace,amplitude=5pt},pState] (3.3,-2.05) -- (11.1,-2.05);
  \node[anchor=north,align=center,font=\scriptsize\bfseries,pState!130!black,text width=9.0cm]
        at (7.2,-2.30)
        {openness without cross-layer governance:\\ the openwashing zone of \citet{widder2024open}};
\end{tikzpicture}
\caption{The openness spectrum of the model layer. Open weights are a necessary but
insufficient condition for a model commons: between gated release and full
commons governance lies the zone in which a model is nominally open yet leaves the
data, compute, and documentation that produced it enclosed, which \citet{widder2024open}
identify as the substance behind much nominal openness. Only the right end combines open
weights with collective stewardship of the upstream layers.}
\label{fig:openness}
\end{figure}
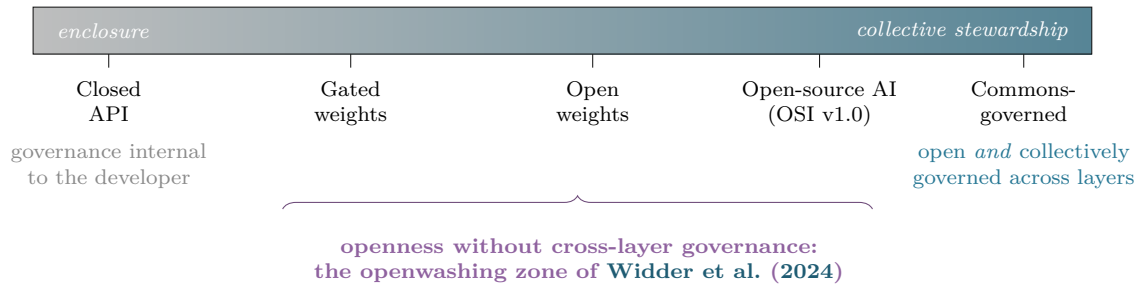

Figure~\ref{fig:openness} arranges the layer along that spectrum. At the closed end a
model is served through an API that exposes behavior but not weights, and governance is
wholly internal to the developer. A gated release discloses weights to approved parties
under contractual terms, a club good in the sense of Section~\ref{sec:taxonomy}. An
open-weights release publishes the parameters for anyone to run and adapt, which is
genuine open access but, by itself, says nothing about who governs the data, the
compute, or the documentation that produced them. The Open Source Initiative's Open
Source AI Definition raises the bar by requiring that the freedoms to use, study,
modify, and share extend to the components needed to exercise them, including
sufficient information about training data \citep{osi2024osaid}, and only at the far end
does a model become commons-governed in the full sense, openly licensed and also
collectively stewarded across the layers that produced it.

The instruments that would let a community govern this layer rather than merely consume
its outputs are transparency and documentation. The Foundation Model Transparency Index
of \citet{bommasani2023foundation} and its 2024 successor \citet{bommasani2024foundation}
score developers across upstream resources including data, data labor, and compute, and
the empirically telling result is that developers score worst precisely on those
upstream indicators, the commons-relevant layers of the stack, even when they score
well on downstream access. Model cards \citep{mitchell2019model} supply the documentation
primitive that turns a released model into an accountable one by recording intended use,
training data, evaluation, and known limitations, and they are the model-layer analogue
of the monitoring principle. The collaborative exemplar of the layer is the BigScience
workshop's BLOOM, a 176-billion-parameter multilingual model built by an open scientific
collaboration on a publicly granted national supercomputer and released under a
Responsible AI License \citep{bigscience2022bloom}. BLOOM instantiates a
compute-commons-plus-open-weights pathway, and paired with the life-cycle carbon
assessment of the same artifact by \citet{luccioni2023estimating}, it demonstrates that a
single model can be collectively built, openly licensed, and transparently measured at
once, which is the cross-layer combination Section~\ref{sec:synthesis} identifies as
durable. In taxonomic terms BLOOM is strong on boundary definition, collective choice,
and monitoring, and it is the clearest empirical point near the open-weight-collaboration
archetype.

The critical counterweight, and the reason the openness spectrum is a secondary axis
rather than the primary one, is the argument of \citet{widder2024open} that nominally
open AI systems are frequently closed in substance and that openness alone does not
reduce the concentration of power. A model released with open weights but trained on
undisclosed data using compute available only to its developer has redistributed an
artifact without redistributing the capacity to produce or contest it. The position paper
of \citet{kapoor2024societal} on the societal impact of open foundation models gives the
most careful benefit-and-risk accounting and argues that the marginal-risk evidence for
the most cited misuse vectors of open weights remains thin, which the taxonomy uses to
resist both a simple equation of openness with safety and its opposite. The combined
lesson of the model layer is the article's organizing constraint: a model commons is not
constituted by open weights, but by open weights conjoined with collective governance of
the data that trained them, the compute that produced them, the documentation that
describes them, and the energy that powered them. Openness is the entry condition; the
commons is the conjunction.

%% file: sec_knowledge.tex
\section{The Knowledge and Evaluation Layer}
\label{sec:knowledge}

The knowledge and evaluation layer is the most fully explained by commons-based peer
production and the least often recognized as a governance object. It comprises the
benchmarks and leaderboards that define what counts as progress, the documentation
standards that make models legible, and the open software toolchains, libraries,
model hubs, and training frameworks on which the entire field runs. These resources
are non-rival in use, anyone may download a benchmark or a library without diminishing
it for others, yet they are subtractable in maintenance, because an unmaintained
benchmark becomes contaminated as models are trained on its test data, and an
unmaintained library decays as the ecosystem around it shifts. The governance question
of the layer is therefore who sets the standard of progress and who bears the upkeep,
and it is answered overwhelmingly by the peer-production dynamics that
\citet{benkler2002coase} first analyzed: loosely coordinated contributors, motivated by
reputation, use value, and reciprocity rather than by price or command, produce and
maintain resources that neither a firm nor a regulator could efficiently supply.

The durability of such commons is not automatic, and the large-N study of free and
open-source software by \citet{schweik2012internet} supplies the evidence base for
which configurations succeed and which are abandoned, identifying the institutional
features, clear governance, low barriers to contribution, and active maintenance, that
separate enduring open-source commons from stalled ones. Those features are Ostrom's
design principles in the software setting, and they transfer directly to the AI
knowledge layer: a benchmark with clear contribution rules, accountable maintainers,
and a mechanism for retiring contaminated items is governed in the sense the taxonomy
intends, while one published once and never curated is mere open access that degrades.
The documentation standards of the layer, model cards \citep{mitchell2019model} for
models and the FAIR principles \citep{wilkinson2016fair} for data and artifacts, are
the monitoring instruments that make the rest of the stack auditable, and their
adoption is itself a collective-choice act of the research community rather than a
regulatory imposition.

This layer is strong on exactly the functions the others are weak on, and weak on the
ones they hold. It excels at provision through contribution and at nested polycentric
organization, since open-source ecosystems are governed in layers from individual
repositories through foundations to cross-project standards bodies, which is Ostrom's
eighth principle realized at scale. It is comparatively weak on graduated sanctions and
on benefit distribution, because peer-production communities have few instruments to
sanction free-riding beyond social exclusion and because the value created by a widely
used benchmark or library is captured disproportionately by the well-resourced actors
best able to exploit it. The layer therefore plays a structural role in the taxonomy
out of proportion to its visibility: it supplies the monitoring and standard-setting
that the data, compute, model, and energy commons need in order to be accountable, and
its peer-production engine is the production process, distinct from the governance of
the resource, that the taxonomy was careful in Section~\ref{sec:commons} to separate.
A mature AI commons does not merely pool data, compute, and weights; it is embedded in a
knowledge commons that tells it how to measure itself, and the health of that knowledge
layer is a precondition for the accountability of all the others.

%% file: sec_energy.tex
\section{The Energy Layer}
\label{sec:energy}

\begin{figure}[t]
\centering
\begin{minipage}[t]{0.49\textwidth}
\centering
\begin{tikzpicture}
\begin{axis}[
   width=0.95\linewidth, height=5.2cm,
   ybar, bar width=14pt,
   ymin=0, ymax=620,
   ylabel={\footnotesize training emissions (t\,CO$_2$e)},
   symbolic x coords={GPT-3, BLOOM (dyn.), BLOOM (life)},
   xtick=data, x tick label style={font=\scriptsize,rotate=18,anchor=east},
   ytick={0,100,200,300,400,500,600},
   tick label style={font=\scriptsize},
   enlarge x limits=0.22,
   nodes near coords, nodes near coords style={font=\scriptsize},
   axis lines=left,
]
\addplot[fill=pGold!75,draw=pGold!120!black] coordinates
   {(GPT-3,552) (BLOOM (dyn.),24.7) (BLOOM (life),50.5)};
\end{axis}
\end{tikzpicture}
\\[-2pt]
{\scriptsize (a) Per-model training footprint.}
\end{minipage}\hfill
\begin{minipage}[t]{0.49\textwidth}
\centering
\begin{tikzpicture}
\begin{axis}[
   width=0.9\linewidth, height=5.2cm,
   ybar, bar width=20pt,
   ymin=0, ymax=1050,
   ylabel={\footnotesize data-centre electricity (TWh)},
   symbolic x coords={2024, 2030 (base)},
   xtick=data, x tick label style={font=\scriptsize},
   ytick={0,200,400,600,800,1000},
   tick label style={font=\scriptsize},
   enlarge x limits=0.5,
   nodes near coords, nodes near coords style={font=\scriptsize},
   axis lines=left,
]
\addplot[fill=pTeal!70,draw=pTeal!130!black] coordinates
   {(2024,415) (2030 (base),945)};
\end{axis}
\end{tikzpicture}
\\[-2pt]
{\scriptsize (b) Global data-centre demand, IEA base case.}
\end{minipage}
\caption{The energy layer in two registers. Panel (a): the training-time carbon
footprint of GPT-3, at roughly $552$~t\,CO$_2$e \citep{patterson2021carbon}, against
the dynamic and full-life-cycle footprints of the openly built BLOOM model, at roughly
$24.7$ and $50.5$~t\,CO$_2$e respectively \citep{luccioni2023estimating}; the
commons-built artifact is the one whose footprint was openly measured. Panel (b): global
data-centre electricity consumption, about $415$~TWh and $1.5\%$ of world electricity in
2024, projected to roughly $945$~TWh and $3\%$ by 2030 under the base case of
\citet{iea2025energyai}. At this scale, siting, grid impact, and carbon intensity become
shared-resource conflicts of the kind community-energy institutions already mediate.}
\label{fig:energy}
\end{figure}
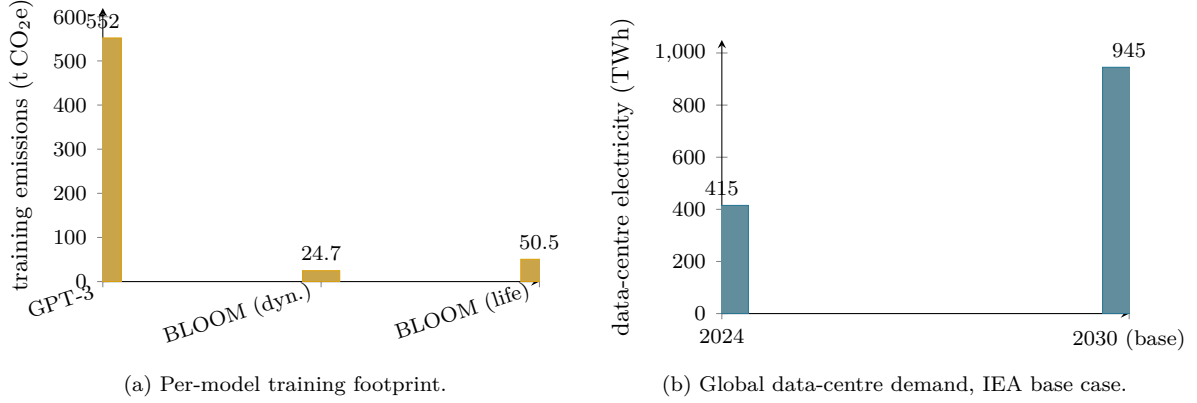

The energy layer is the physical substrate of the others, and treating it as a
first-class commons rather than as an externality to be priced is one of the
article's organizing choices. The case for doing so rests first on the measurement
literature that made the cost of computation legible, and second on the macro-scale
demand projections that turn that cost into a governance problem at the level of
national grids. \citet{strubell2019energy} performed the first widely cited
accounting of the financial and carbon cost of training natural-language models and
drew the distributional conclusion that compute-intensive research enacts a choice
about who can afford to participate, a conclusion that ties the energy layer directly
to the compute divide of Section~\ref{sec:compute}. \citet{patterson2021carbon}
refined the per-model accounting across a range of large models and showed that the
choices of model architecture, data-center location, and processor can swing the
carbon footprint by roughly two to three orders of magnitude, which converts
sustainability from a fixed externality into a governable design variable, and they
estimated the training of GPT-3 at on the order of $552$ tonnes of carbon-dioxide
equivalent. \citet{luccioni2023estimating} then extended the boundary of measurement
to a full life cycle for the 176-billion-parameter BLOOM model, reporting
approximately $24.7$ tonnes of carbon-dioxide equivalent from the dynamic power of
final training and approximately $50.5$ tonnes once equipment manufacturing and
operational overhead are included, and in doing so demonstrated that a collaboratively
and openly built model can also be a transparently measured one. The contrast,
displayed in Figure~\ref{fig:energy}, is the point: the difference between the two
artifacts is not only scale but governance, since the model built as a commons was the
model whose footprint was openly accounted for.

The macro-scale of the layer is what makes its governance unavoidable.
\citet{masanet2020recalibrating} recalibrated global data-center energy estimates
downward by accounting for the efficiency gains of the preceding decade, supplying the
rigorous baseline against which AI-driven growth is now measured, and the International
Energy Agency's analysis of energy and AI reports that data centers accounted for about
$1.5\%$ of world electricity consumption in 2024 and projects that share to roughly
double by 2030 under its base case, driven disproportionately by AI-accelerated servers
\citep{iea2025energyai}. At this scale the appropriation of grid capacity, the siting of
facilities, and their water draw for cooling become precisely the shared-resource
conflicts that community-energy institutions have long mediated. The taxonomy makes the
analogy concrete through two strands. The carbon-aware scheduling of
\citet{radovanovic2023carbon} shifts temporally flexible data-center load toward
low-carbon hours, which is an appropriation rule in Ostrom's sense, congruence between
the timing of withdrawal and the condition of the shared grid, implemented in software.
The study of energy-community surplus sharing by \citet{wade2025energy} examines how
community-energy cooperatives distribute surplus under solidarity rather than pure-market
logic, and it supplies the cleanest existing template for an energy-commons cell, strong
on collective choice and on benefit distribution to members.

The energy layer therefore does two things in the taxonomy. It establishes a provision
obligation that we elevate to a secondary attribute of every institution: any commons
that consumes compute consumes energy, and the monitoring principle, applied to the
energy layer, requires that this consumption be accounted for, so that a compute or
model commons that does not measure and govern its energy footprint is incompletely
governed by its own standard. And it connects AI governance to an established commons
tradition with its own mature institutional forms, the energy cooperative and the
community-energy scheme, from which the AI commons can borrow not only the analogy but
the governance machinery. The efficiency-as-inclusivity argument of the Green-AI program
\citep{schwartz2020green} and the systematic-reporting framework of
\citet{henderson2020towards} together supply the tooling, lower-footprint methods and
standardized carbon accounting, through which a commons can meet that provision
obligation, closing the loop between the measurement literature that opens the section
and the governance project that the article as a whole advances.

%% file: sec_synthesis.tex
\section{Synthesis: The Taxonomy Applied}
\label{sec:synthesis}

The five layer sections recur to the same institutional forms recombined over
different resources, and we now assemble those forms into the ten archetypes that
serve as landmarks in the classificatory space of Section~\ref{sec:taxonomy}. The
archetypes are not the classes of the taxonomy, which is polythetic, but recurrent
regions near which real institutions cluster; Table~\ref{tab:archetypes} records, for
each, the resource layers it pools, the legal vehicle through which it holds rights,
its position on the openness spectrum, its rule for benefit distribution, its
sustainability stance, the Ostromian functions on which it is strong, and its
empirical exemplars. Reading down the table recovers the article's central claim. The
archetypes that pool a single layer, the bare data cooperative, the bare open-weight
release, the bare public compute pool, are each strong on a characteristic subset of
functions and weak on the rest, and their weaknesses are systematically the functions
held by other layers: the data cooperative lacks compute, the open-weight release
lacks data and compute governance, the public compute pool lacks user collective
choice. The archetypes that endure, by contrast, are cross-layer conjunctions, the
federated consortium that pools data governance and compute, the open-weight
collaboration that pools compute, model, and documentation, the public-AI initiative
that reaches across all five, and they are durable precisely because they supply each
other's missing functions.

\begin{sidewaystable}
\centering
\caption{Ten archetypes of commons-governed AI located in the taxonomy. Layers:
\stack{D} data, \stack{C} compute, \stack{M} models, \stack{K} knowledge/evaluation,
\stack{E} energy. Functions (strong): B boundaries, A appropriation/provision, C
collective choice, M monitoring, S sanctions, R conflict resolution, L recognition,
N nesting. Openness is read on the spectrum of Figure~\ref{fig:openness}.}
\label{tab:archetypes}
\footnotesize
\begin{tabularx}{\textheight}{@{}p{2.9cm} c p{2.0cm} p{1.9cm} p{2.4cm} p{2.0cm} p{1.5cm} X@{}}
\toprule
\textbf{Archetype} & \textbf{Layers} & \makecell[l]{\textbf{Legal}\\\textbf{vehicle}} & \textbf{Openness} &
\textbf{Benefit} & \textbf{Sustainab.} & \textbf{Strong} & \textbf{Exemplars} \\
\midrule
Data trust & D & Trust (fiduciary) & Club / open & Beneficiaries & Indirect &
B, L, R & \citet{delacroix2019bottomup} \\
Data cooperative & D & Cooperative & Club & Members & Indirect &
A, C, L & \citet{hardjono2019datacoops} \\
Open dataset commons & D, K & License + community & Open & Public & Reported &
A, B, N & Common Voice \citep{ardila2020commonvoice} \\
Indigenous data sovereignty & D & Community / customary & Restricted by authority &
Community & Indirect & C, B & CARE \citep{carroll2020care}; \citet{kukutai2016indigenous} \\
Federated learning consortium & D, C & Consortium agreement & Club & Members & Partial &
B, M, C & \citet{mcmahan2017communication}; \citet{kairouz2021advances} \\
Public compute commons & C, E & Public agency / JU & Open (eligibility) & Public &
Governable & M, L, N & NAIRR \citep{nairr2023strengthening}; EuroHPC \citep{eurohpc2024aifactories} \\
Open-weight model collab.\ & C, M, K & Foundation + RAIL & Open weights & Public & Measured &
B, C, M & BLOOM \citep{bigscience2022bloom,luccioni2023estimating} \\
Knowledge / FLOSS commons & K & Foundation / project & Open & Public & Reported &
A, N, M & \citet{schweik2012internet}; model cards \citep{mitchell2019model} \\
Energy / carbon-aware coop.\ & E, C & Cooperative / scheme & Open & Members + grid &
Core & C, A & \citet{wade2025energy}; \citet{radovanovic2023carbon} \\
Public AI / DPI & D, C, M, K, E & Public + DAO option & Open + governed & Public &
Core & C, L, N & \citet{publicai2024whitepaper}; \citet{hassan2021dao} \\
\bottomrule
\end{tabularx}
\end{sidewaystable}

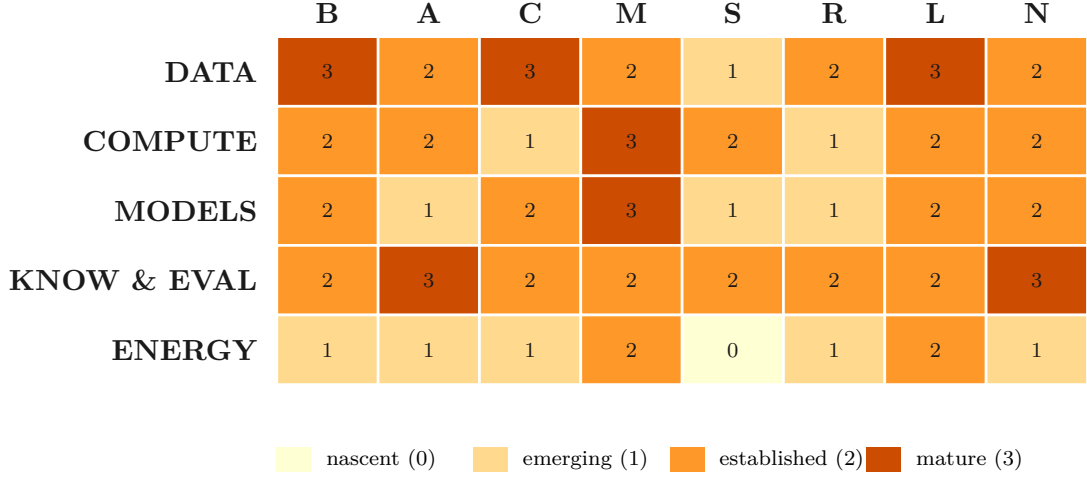
\begin{figure}[t]
\centering
\begin{tikzpicture}[font=\small]
  \def\dx{1.34}\def\dy{0.92}
  \definecolor{Y0}{HTML}{FFFFD4}
  \definecolor{Y1}{HTML}{FED98E}
  \definecolor{Y2}{HTML}{FE9929}
  \definecolor{Y3}{HTML}{CC4C02}
  \foreach \j/\lab in {0/B,1/A,2/C,3/M,4/S,5/R,6/L,7/N}{
     \node[font=\bfseries,pInk] at (\j*\dx+\dx/2,5*\dy+0.32){\lab};}
  \foreach \r/\lab in {4/DATA,3/COMPUTE,2/MODELS,1/{KNOW \& EVAL},0/ENERGY}{
     \node[anchor=east,font=\bfseries,pInk] at (-0.12,\r*\dy+\dy/2){\lab};}
  \foreach \ry/\j/\val in {%
     4/0/3,4/1/2,4/2/3,4/3/2,4/4/1,4/5/2,4/6/3,4/7/2,%
     3/0/2,3/1/2,3/2/1,3/3/3,3/4/2,3/5/1,3/6/2,3/7/2,%
     2/0/2,2/1/1,2/2/2,2/3/3,2/4/1,2/5/1,2/6/2,2/7/2,%
     1/0/2,1/1/3,1/2/2,1/3/2,1/4/2,1/5/2,1/6/2,1/7/3,%
     0/0/1,0/1/1,0/2/1,0/3/2,0/4/0,0/5/1,0/6/2,0/7/1}{
       \fill[Y\val] (\j*\dx,\ry*\dy) rectangle (\j*\dx+\dx,\ry*\dy+\dy);
       \draw[white,line width=1pt] (\j*\dx,\ry*\dy) rectangle (\j*\dx+\dx,\ry*\dy+\dy);
       \node[font=\scriptsize,pInk] at (\j*\dx+\dx/2,\ry*\dy+\dy/2){\val};
  }
  \begin{scope}[shift={(0,-1.15)}]
    \foreach \v/\lab in {0/nascent,1/emerging,2/established,3/mature}{
       \fill[Y\v] (\v*2.6,0) rectangle (\v*2.6+0.45,0.32);
       \node[anchor=west,font=\scriptsize] at (\v*2.6+0.52,0.16){\lab\ (\v)};}
  \end{scope}
\end{tikzpicture}
\caption{Maturity matrix of commons-governed AI. Cell $(\,\text{layer},\,\text{function})$
records the authors' synthesized assessment of how developed the field's institutions
are on a four-point scale (sequential YlOrBr; darker is more mature). The data row is the
most uniformly developed; monitoring is strong on compute and models but emerging on
energy; the graduated-sanctions column and the energy row are the thinnest regions, the
two clearest frontiers for institutional design. The assessment maps the evidence of
Sections~\ref{sec:data}--\ref{sec:energy}; it is a map, not a measurement.}
\label{fig:heatmap}
\end{figure}

The coverage of the classificatory space is uneven, and Figure~\ref{fig:heatmap}
records our assessment of how mature the field's institutions are at each (layer,
function) cell on a four-point scale from nascent to mature. The assessment is
qualitative and synthesizes the evidence of Sections~\ref{sec:data}--\ref{sec:energy};
it is offered as a map of where institutional development is concentrated and where it
is thin, not as a measurement. Three patterns are legible. The data row is the most
uniformly developed, because the trust, the cooperative, and the sovereignty regime
together cover boundaries, collective choice, and recognition with established legal
vehicles. The monitoring column is strong across the compute and model rows, because
compute is intrinsically measurable \citep{sastry2024computing} and because
transparency instruments exist for models \citep{bommasani2024foundation,mitchell2019model},
but it is only emerging on the energy row, where carbon accounting is recent. The
graduated-sanctions column and the energy row are the thinnest regions of the entire
space, which identifies the two clearest frontiers for institutional design: the AI
commons has few instruments to sanction free-riding short of exclusion, and it has
barely begun to govern its own energy footprint as a provision obligation.

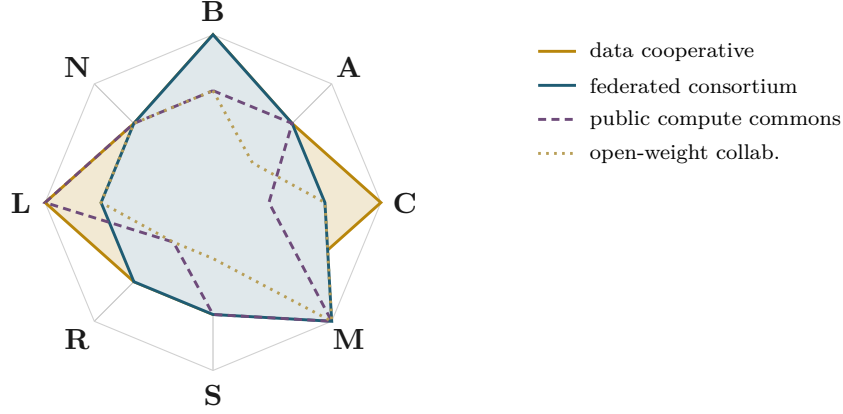
\begin{figure}[t]
\centering
\begin{tikzpicture}[font=\scriptsize,scale=1.0]
  \def\u{0.74} 
  \foreach \ang/\lab in {90/B,45/A,0/C,315/M,270/S,225/R,180/L,135/N}{
     \draw[pMute!55] (0,0) -- (\ang:3*\u);
     \node[font=\bfseries,pInk] at (\ang:3*\u+0.32){\lab};
  }
  \foreach \r in {1,2,3}{
     \draw[pMute!40]
       (90:\r*\u)--(45:\r*\u)--(0:\r*\u)--(315:\r*\u)--(270:\r*\u)--
       (225:\r*\u)--(180:\r*\u)--(135:\r*\u)--cycle;
  }
  \draw[pGold,line width=1.1pt,fill=pGold!18]
    (90:3*\u)--(45:2*\u)--(0:3*\u)--(315:2*\u)--(270:2*\u)--(225:2*\u)--(180:3*\u)--(135:2*\u)--cycle;
  \draw[pTeal,line width=1.1pt,fill=pTeal!14]
    (90:3*\u)--(45:2*\u)--(0:2*\u)--(315:3*\u)--(270:2*\u)--(225:2*\u)--(180:2*\u)--(135:2*\u)--cycle;
  \draw[pState,line width=1.1pt,dash pattern=on 3pt off 2pt]
    (90:2*\u)--(45:2*\u)--(0:1*\u)--(315:3*\u)--(270:2*\u)--(225:1*\u)--(180:3*\u)--(135:2*\u)--cycle;
  \draw[pGoldL!80!black,line width=1.1pt,dash pattern=on 1pt off 2pt]
    (90:2*\u)--(45:1*\u)--(0:2*\u)--(315:3*\u)--(270:1*\u)--(225:1*\u)--(180:2*\u)--(135:2*\u)--cycle;
  \begin{scope}[shift={(4.3,2.0)}]
    \draw[pGold,line width=1.1pt](0,0)--(0.5,0); \node[anchor=west,font=\scriptsize]at(0.55,0){data cooperative};
    \draw[pTeal,line width=1.1pt](0,-0.45)--(0.5,-0.45); \node[anchor=west,font=\scriptsize]at(0.55,-0.45){federated consortium};
    \draw[pState,line width=1.1pt,dash pattern=on 3pt off 2pt](0,-0.90)--(0.5,-0.90); \node[anchor=west,font=\scriptsize]at(0.55,-0.90){public compute commons};
    \draw[pGoldL!80!black,line width=1.1pt,dash pattern=on 1pt off 2pt](0,-1.35)--(0.5,-1.35); \node[anchor=west,font=\scriptsize]at(0.55,-1.35){open-weight collab.};
  \end{scope}
\end{tikzpicture}
\caption{Four archetypes profiled against the eight design principles
(B boundaries, A appropriation/provision, C collective choice, M monitoring,
S sanctions, R conflict resolution, L recognition, N nesting), on the four-point
maturity scale of Figure~\ref{fig:heatmap}. No single archetype fills the octagon: the
data cooperative is strong on collective choice and recognition, the public compute
commons on monitoring and nesting, and the two others interpolate. A commons strong on
every principle is necessarily a nested conjunction across layers, which is Ostrom's
eighth principle and the article's central design claim.}
\label{fig:radar}
\end{figure}

A complementary reading compares specific archetypes against the eight design
principles directly. Figure~\ref{fig:radar} profiles four archetypes that pool
overlapping layers, the data cooperative, the federated health consortium, the public
compute commons, and the open-weight model collaboration, on the same four-point scale.
The profiles make the complementarity visible. The data cooperative is strong where the
public compute commons is weak, on collective choice and benefit distribution, and the
public compute commons is strong where the cooperative is weak, on monitoring,
recognition, and nesting; the federated consortium and the open-weight collaboration
interpolate between them. No single archetype fills the octagon, which is the
visual statement of the article's thesis: a commons that scores high on every design
principle is necessarily cross-layer, because the principles are distributed across the
layers and no single-layer institution holds them all. The durable AI commons is a
nested conjunction of these archetypes, an instance of Ostrom's eighth principle, in
which a data cooperative supplies collective choice, a public compute pool supplies
monitorable infrastructure, an open-weight collaboration supplies the model and its
documentation, and an energy scheme supplies the sustainability accounting, each nested
within a polycentric whole.

The synthesis yields one observation that organizes the applications and tensions of
the next two sections.

\begin{observation}[Cross-layer durability]
A commons-governed AI institution that pools a single resource layer is strong only on
the governance functions native to that layer and remains exposed on the others, so
that its outputs can be re-enclosed at any unpooled layer. Durable commons are
cross-layer conjunctions in which each pooled layer supplies functions the others lack,
organized as nested enterprises in the sense of Ostrom's eighth design principle.
\end{observation}

\noindent This is the lens through which the sectoral instances of
Section~\ref{sec:applications} are read, and the standard against which the failure
modes of Section~\ref{sec:tensions} are diagnosed.

%% file: sec_applications.tex
\section{Sectoral Applications}
\label{sec:applications}

The cross-layer durability observation of Section~\ref{sec:synthesis} is most clearly
illustrated by the sectors in which commons-governed AI is already operating, because
each sector forces a particular conjunction of layers and reveals which functions must
be supplied together. We read four.

Health is the sector in which the data layer cannot be pooled directly, and it is
therefore the natural home of the federated learning consortium. Patient data are
bound to the institution that holds them by regulation, by sovereignty, and by ethics,
so a commons of raw clinical data is foreclosed; what can be pooled is learning, through
the federated-averaging mechanism of \citet{mcmahan2017communication} and the broader
apparatus surveyed by \citet{kairouz2021advances}, in which each hospital contributes
computation local to its own records and only model updates are aggregated. The
resulting consortium is a cross-layer commons of the data and compute rows that supplies
strong boundary definition and monitoring, because participation and contribution are
explicit and auditable, while its collective-choice strength depends on whether the
member institutions jointly set the training objective. The health case also exhibits
the layer's characteristic frontier, graduated sanctions, since a consortium's only
instrument against a member that contributes corrupted updates is exclusion, which is the
sanction Ostrom's fifth principle warns should be graduated rather than binary.

Science is the sector in which the knowledge layer is primary and the commons is the
default rather than the exception. The research enterprise has long produced its data,
benchmarks, and tools through commons-based peer production \citep{benkler2002coase},
and the stewardship principles of the layer, FAIR for data and artifacts
\citep{wilkinson2016fair} and the documentation primitives of model cards
\citep{mitchell2019model}, originate in scientific practice. The Governing Knowledge
Commons program of \citet{frischmann2014governing} was built substantially on
scientific cases, and the open-weight model collaboration of the model layer is, in its
governance, an extension of scientific peer production to artifacts that were previously
proprietary, as the BigScience workshop's BLOOM demonstrates \citep{bigscience2022bloom}.
Science is therefore the sector that supplies the AI commons with its production engine
and its monitoring norms, and the one in which the knowledge layer's strength on
provision and nesting is most fully realized.

Public-interest and sovereign AI is the sector in which the compute and energy layers
become the binding constraint and the state enters as a commons participant rather than
as a regulator. The public-AI program \citep{publicai2024whitepaper} and the national
compute initiatives of Section~\ref{sec:compute} \citep{nairr2023strengthening,
eurohpc2024aifactories} respond to the compute divide by pooling publicly funded
infrastructure, and their aspiration, only partly realized, is to reach across all five
layers, pooling public data, public compute, openly licensed models, open toolchains,
and accountable energy. The taxonomy locates these initiatives precisely by what they
have not yet pooled, user collective choice, and the agenda of Section~\ref{sec:agenda}
treats the closing of that gap as the central design task of the sector. The
decentralized-governance mechanism of \citet{hassan2021dao} appears here as one
candidate instrument for supplying the collective choice that the administrative form
lacks, though its own governance pathologies, addressed in the next section, make it a
candidate rather than a solution.

Language and cultural revitalization is the sector in which the data layer's collective
and community dimension is decisive, and in which open access and commons governance can
directly conflict. The CARE principles \citep{carroll2020care} and the Indigenous
data-sovereignty movement \citep{kukutai2016indigenous} establish that data about a
community are governed by that community, which means that a language commons may
legitimately restrict appropriation to honor collective authority even as it pools
provision through voluntary contribution, as the multilingual Common Voice corpus does
in its consent-based design \citep{ardila2020commonvoice}. This sector is the clearest
refutation of the equation of commons with open access: it is governed precisely by the
distinction Section~\ref{sec:commons} draws, and it supplies the normative content that
the more infrastructural sectors lack. Across all four, the pattern of
Section~\ref{sec:synthesis} holds, that the sector determines which layers must be
pooled together, and that the durable institution is the one that supplies, across those
layers, the functions no single layer holds alone.

%% file: sec_tensions.tex
\section{Tensions, Failure Modes, and Trade-offs}
\label{sec:tensions}

A taxonomy that only catalogued successes would be an advertisement rather than an
analysis. The commons frame inherits from Ostrom not only the design principles but the
recognition that commons fail when those principles are violated, and the AI setting
introduces failure modes of its own. We name five, each diagnosed through the
cross-layer lens of Section~\ref{sec:synthesis}.

The first and most important is openwashing, the appropriation of the language of
openness by arrangements that leave the underlying concentration of power intact.
\citet{widder2024open} argue that nominally open AI systems are frequently closed in
substance, because a model released with open weights but trained on undisclosed data
using compute available only to its developer has redistributed an artifact without
redistributing the capacity to produce or contest it. This diagnosis has an
intellectual precedent in the critique of permissive openness advanced by
\citet{kleiner2010telekommunist}, who argued that artifacts and licenses which are
open without restricting the asymmetric accumulation they permit are readily absorbed
by capital, and whose proposed remedy of reciprocity-conditioned, copyfarleft
licensing anticipates the share-alike instruments examined below. In the taxonomy this is the
single-layer fragility of the cross-layer durability observation made acute: openness at
the model layer alone, decoupled from governance of the data, compute, and energy layers,
is not a commons but a marketing posture, and it is the zone marked in
Figure~\ref{fig:openness}. The defense is structural rather than rhetorical, the
insistence that a model commons be scored on the upstream indicators where the
Foundation Model Transparency Index shows developers scoring worst
\citep{bommasani2024foundation}, and that openness be treated as the entry condition to
the commons rather than as its achievement.

The second is the compute bottleneck. The properties that make compute the most
governable layer, its excludability and concentration \citep{sastry2024computing}, make
it the layer at which an otherwise well-governed commons is most easily strangled, since
a data cooperative, an open-weight collaboration, and a knowledge commons can each be
exemplary and still depend on frontier compute that only a few actors command. The public
compute initiatives of Section~\ref{sec:compute} are the structural response, but their
distance from user collective choice means that the bottleneck can be relieved without
being democratized, and a compute commons that pools provision while concentrating
decision rights reproduces the very asymmetry it was meant to dissolve. This is the
trade-off between the second and third design principles, provision and collective
choice, in its sharpest form.

The third is free-riding, the classical commons pathology, which the AI setting
sharpens because the non-rivalry of the informational layers makes appropriation
costless while provision remains expensive. A widely used benchmark, library, or open
model is consumed disproportionately by the well-resourced actors best able to exploit
it, who contribute least to its maintenance, and the knowledge layer's characteristic
weakness on graduated sanctions, the thinnest column of Figure~\ref{fig:heatmap}, means
the commons has few instruments short of social exclusion to correct it. Reciprocal and
share-alike licensing, the Responsible AI License under which BLOOM was released
\citep{bigscience2022bloom} being one instance, is the architectural attempt to convert
appropriation into provision, and its enforceability is among the open problems of
Section~\ref{sec:agenda}.

The fourth is the tension between scale and sustainability. The energy accounting of
Section~\ref{sec:energy} shows that capability has been bought partly with carbon
\citep{patterson2021carbon,iea2025energyai}, and a commons that competes with enclosed
actors on the frontier inherits their footprint, so that the provision obligation we
attached to the energy layer can conflict with the ambition to match proprietary
capability. The Green-AI argument that efficiency is itself a form of inclusivity
\citep{schwartz2020green} reframes the tension as a design choice rather than a dilemma,
but it does not dissolve it: a commons may have to choose between matching frontier scale
and honoring its own sustainability stance, and the taxonomy makes that choice visible by
recording the sustainability stance as a first-class attribute rather than burying it.

The fifth is governance overhead and the pathologies of the vehicles themselves. The
collective-choice machinery that distinguishes a commons from a firm is costly, and the
decentralized-autonomous-organization form proposed as one instrument for it
\citep{hassan2021dao} carries its own failure modes, plutocratic capture by
token-weighted voting, low participation, and the substitution of code for accountable
judgment, the last being the dark side of the architecture-as-law insight of
\citet{lessig2006code}. The trust and cooperative forms are more robust but slower, and
the standing critique that principles alone cannot govern \citep{mittelstadt2019principles}
applies with equal force to institutional forms adopted as slogans rather than operated
with fidelity to the design principles. Across all five failure modes the diagnosis is
the same: the commons fails where it is single-layer, where provision and collective
choice are decoupled, or where a vehicle is adopted without the rules-in-use that make
it work, and the remedy is the disciplined cross-layer, polycentric construction that the
agenda now sets out.

%% file: sec_agenda.tex
\section{A Research and Policy Agenda}
\label{sec:agenda}

The taxonomy is a map, and a map is useful chiefly for the routes it reveals. The
maturity matrix of Figure~\ref{fig:heatmap} and the failure modes of
Section~\ref{sec:tensions} together identify where the institutional development of
commons-governed AI is thin and where it is contested, and we close by setting out the
research and policy agenda those gaps imply, organized by the design principle each
addresses.

The thinnest column of the matrix is graduated sanctions, and the first item on the
agenda is the design of enforcement instruments for the AI commons that are graduated
rather than binary. The only sanction most current arrangements possess is exclusion,
which Ostrom's fifth principle warns is too blunt to sustain cooperation, and the open
research question is how reciprocal and share-alike licensing, contribution accounting,
and reputation systems can be combined into a graduated response to free-riding that is
enforceable across the non-rival informational layers. The enforceability of the
Responsible AI License family \citep{bigscience2022bloom} and the design of
contribution-weighted governance are concrete sub-problems.

The thinnest row of the matrix is energy, and the second item is to make the provision
obligation we attached to that layer operational. This requires standardized, audited
carbon and water accounting as a condition of participation in any compute-bearing
commons, building on the systematic-reporting framework of \citet{henderson2020towards}
and the life-cycle methodology of \citet{luccioni2023estimating}, and it requires
importing the governance machinery of community-energy commons \citep{wade2025energy} and
carbon-aware scheduling \citep{radovanovic2023carbon} into the operation of AI compute
pools. The policy lever is to attach sustainability-accounting requirements to public
compute provision \citep{nairr2023strengthening,eurohpc2024aifactories}, where the state
already sits inside the commons and can set provision rules.

The most contested cell is the conjunction of the compute layer with collective choice,
and the third item is to close the distance between public compute provision and user
self-governance. The public initiatives pool provision but set allocation
administratively, and the research question is what collective-choice mechanism, an
elected allocation board, a federated council of user institutions, or a carefully
bounded decentralized-governance layer \citep{hassan2021dao} that avoids the plutocratic
capture diagnosed in Section~\ref{sec:tensions}, can vest allocation in the user community
without sacrificing the accountability the public form provides. This is the central
design task of the public-AI sector.

\begin{figure}[t]
\centering
\begin{tikzpicture}[font=\small]
  \definecolor{N1}{HTML}{1F5C73}\definecolor{N2}{HTML}{3E7C8C}
  \definecolor{N3}{HTML}{6FA8B8}\definecolor{N4}{HTML}{B8860B}
  \draw[N1,line width=1.4pt,fill=N1!7,rounded corners=14pt] (0,0) rectangle (12.4,6.4);
  \node[anchor=north west,font=\footnotesize\bfseries,N1] at (0.45,6.18) {Ecosystem: cross-project standards \& public infrastructure};
  \draw[N2,line width=1.2pt,fill=N2!9,rounded corners=12pt] (0.7,0.6) rectangle (11.7,5.3);
  \node[anchor=north west,font=\footnotesize\bfseries,N2] at (1.05,5.08) {Foundation: trusts, cooperatives, consortia};
  \draw[N3,line width=1.1pt,fill=N3!12,rounded corners=10pt] (1.5,1.2) rectangle (10.9,4.2);
  \node[anchor=north west,font=\footnotesize\bfseries,N3!130!black] at (1.85,3.98) {Project: a corpus, a model, a benchmark, a pool};
  \draw[N4,line width=1.1pt,fill=N4!15,rounded corners=8pt] (2.4,1.7) rectangle (10.0,3.2);
  \node[font=\footnotesize\bfseries,N4!130!black] at (6.2,2.78) {Contributor / member};
  \node[font=\scriptsize,pInk,align=center] at (6.2,2.25)
        {individual rules-in-use: contribution, appropriation,\\ monitoring, local conflict resolution};
  \node[anchor=north west,font=\scriptsize,pMute,align=left,text width=12.0cm] at (0.15,-0.35)
     {Each layer supplies the governance functions the layer within it lacks; the whole is a nested enterprise (Ostrom's eighth principle).};
\end{tikzpicture}
\caption{A polycentric architecture for a durable AI commons. Governance is organized in
nested layers, from the individual contributor through the project that pools a single
resource, the foundation or cooperative that holds rights and supplies collective choice,
to the ecosystem layer of standards and public infrastructure. The cross-layer durability
observation of Section~\ref{sec:synthesis} is realized vertically: each enclosing layer
supplies the functions the enclosed layer cannot hold alone, which is the
institutional content of Ostrom's eighth design principle applied to the AI stack.}
\label{fig:nesting}
\end{figure}

Underlying these three specific items is a single structural prescription, drawn from
the cross-layer durability observation and depicted in Figure~\ref{fig:nesting}: the
durable AI commons is polycentric and nested. No single institution will hold every
design principle over every layer, and the attempt to build one would reproduce the
monocentric concentration the commons exists to dissolve. The architecture the evidence
supports instead is a nesting of enterprises, a contributor within a project, a project
within a foundation or cooperative, a foundation within an ecosystem of standards and
public infrastructure, in which each enclosing layer supplies the governance functions
the layer within it cannot hold alone. This is Ostrom's eighth principle restated for the
AI stack, and it is the form in which the data cooperative, the federated consortium, the
public compute commons, the open-weight collaboration, and the energy scheme of the
preceding sections compose into a single governable whole rather than competing as rival
proposals.

The agenda has a methodological coda. The taxonomy is offered as a living instrument, and
its maturity matrix is a snapshot of a fast-moving field; the responsible use of it is to
re-examine the cells before each application, because the thin regions of
Figure~\ref{fig:heatmap} are precisely where institutional innovation is most active and
where the map will date fastest. The taxonomy's value is not the particular shading of its
cells but the axes that generate them, the five layers and the eight functions, which are
inherited from a validated institutional grammar and will outlast any single snapshot of
the practice they classify.

%% file: sec_conclusion.tex
\section{Conclusion}
\label{sec:conclusion}

The governance of artificial intelligence has been theorized as a choice between the
market and the state, between private control of the AI stack and its regulation from
above. This article has argued that a third institutional family is already operating in
the gap between them, and that it forms a coherent object of study once it is named.
Commons-governed artificial intelligence is the collective, self-organized stewardship of
the resources of the AI stack by the communities that produce and depend on them, and the
analytic tradition built by Ostrom and her successors for common-pool and knowledge
commons is the right instrument for classifying it. We have proposed a two-dimensional
taxonomy whose axes are inherited from that tradition rather than constructed ad hoc, the
five resource layers of data, compute, models, knowledge, and energy on the one axis, and
the eight design principles of self-governing institutions on the other, and we have
populated the resulting space by examining the published evidence layer by layer and
locating ten recurrent institutional archetypes within it.

Three findings organize the contribution. The first is that the AI stack is genuinely
commons-like at every layer, because each layer is subtractable in a governance-relevant
dimension even where its informational content is non-rival, so that the vocabulary of
provision, appropriation, monitoring, and sanction applies to corpora, weights, compute,
benchmarks, and energy alike. The second, recorded in the maturity matrix and the
principle-by-principle comparison, is that no single-layer institution holds every design
principle, and that the durable commons are cross-layer conjunctions in which each pooled
layer supplies the functions the others lack, organized as nested polycentric enterprises.
The third is that the energy and carbon cost of computation is a commons-governance
problem on a par with the governance of data and weights, a provision obligation of any
compute-bearing commons rather than an externality, and that AI governance accordingly has
much to learn from the mature institutional forms of the community-energy tradition.

The taxonomy answers a standing challenge. The critique that ethical principles alone
cannot govern AI, that a list of values without fiduciary duties, accountability
mechanisms, and collective-choice arrangements is institutionally empty, is correct, and
it has wanted a constructive reply. The reply is institutional: the trust supplies the
fiduciary duty, the cooperative the collective choice, the federated consortium the
auditable boundary, the public compute pool the monitorable infrastructure, the open-weight
collaboration the documented artifact, and the energy scheme the sustainability accounting,
each nested within a polycentric whole that no single one could constitute. None of these is
sufficient alone, which is the central lesson of the taxonomy, and the openwashing, compute
bottleneck, free-riding, scale-sustainability, and governance-overhead failure modes are
each the signature of a commons that stopped at one layer or adopted a vehicle without its
rules-in-use. The map this article draws is a snapshot of a field moving fast enough that
its cells will need re-examination, but its axes are fixed, and the third way between the
market and the state is, on the evidence assembled here, neither a tragedy to be averted by
enclosure nor a utopia to be proclaimed, but an institutional design problem with a
tractable structure and a growing body of working solutions.